\documentclass[a4paper,11pt,floatfix]{article}
\pdfoutput=1 % if your are submitting a pdflatex (i.e. if you have
\usepackage{jheppub} % for details on the use of the package, please
\usepackage[T1]{fontenc} % if needed

\usepackage{amsmath}
\usepackage{dsfont}
\usepackage{slashed}
\usepackage{bbold}
\usepackage{psfrag}
\PassOptionsToPackage{caption=false}{subfig}
\usepackage{subfig}
\usepackage{multirow}
\usepackage{booktabs}
\usepackage{hyperref}
\usepackage{feynmp}
\usepackage{color}
\usepackage[normalem]{ulem} % allows strikethru

\usepackage[utf8]{inputenc}

\newcommand{\beq}{\begin{equation}}
\newcommand{\eeq}{\end{equation}}
\newcommand{\bea}{\begin{eqnarray}}
\newcommand{\eea}{\end{eqnarray}}

\newcommand{\Ylm}{\mathcal{Y}}
\newcommand{\rv}{\textbf{r}}

\newcommand{\qv}{\textbf{q}}
\newcommand{\qtv}{\tilde{\textbf{q}}}
\newcommand{\qt}{\tilde{q}}
\newcommand{\Ot}{\Omega_{\qt}}

\newcommand{\Kv}{\textbf{K}}

\newcommand{\mum}{\frac{\mu_{12}}{m_1}}

\newcommand{\DE}{\Delta E_B}
\newcommand{\sigmap}{\overline{\sigma}_n}
\newcommand{\sigmaa}{\overline{\sigma}_{\rm nuc}}

\newcommand{\Y}{\mathcal{Y}}
\newcommand{\Fdis}{|F_{\rm dis}(\qv,\tilde{\text{q}})|^2}
\newcommand{\Fdisq}{|F_{\rm dis}(\text{q},\tilde{\text{q}})|^2}

\newcommand{\FdisE}{\ifmmode {\lvert F_{\rm dis}(E_r,E_{\rm int}) \rvert}^2 \else ${\lvert F_{\rm dis}(E_r,E_{\rm int}) \rvert}^2$\fi}
\newcommand{\Fmol}{\ifmmode {\lvert F_{\rm mol}(\mathbf{q,\tilde{\mathbf{q}}}) \rvert}^2 \else ${\lvert F_{\rm mol}(\mathbf{q},\tilde{\mathbf{q}}) \rvert}^2$\fi}
\newcommand{\FDM}{|F_{\rm DM}(\mathbf{q})|^{2}}
\newcommand{\fv}{f(\mathbf{v})}
\newcommand{\Mfi}{\ifmmode {\lvert \mathcal{M}_{2\to2} \rvert}^2 \else ${\lvert \mathcal{M}_{2\to2} \rvert}^2$\fi}

\newcommand{\mdm}{m_{\chi}}

\newcommand{\units}[1]{\mathrm{\; #1}}

\definecolor{Jcolor}{rgb}{0.9,0.3,0.5}

\renewcommand{\subsubsection}[1]{\addtocounter{subsubsection}{1}
\par\nobreak
\medskip
\nobreak
\noindent{\it \thesubsubsection.  #1 }
\par\nobreak\medskip\nobreak}

\def\lpar#1#2#3#4{\rlap{\raise#3\hbox{$\hskip#4#1\left\{\mbox{\phantom{\rule[0mm]{0mm}{#2}}}\right.$}}}
\def\rpar#1#2#3#4{\rlap{\raise#3\hbox{$\hskip#4\left\}#1\mbox{\phantom{\rule[0mm]{0mm}{#2}}}\right.$}}}

\title{\boldmath Detection of sub-GeV Dark Matter and Solar Neutrinos via Chemical-Bond Breaking}

\author[a]{Rouven Essig,}
\emailAdd{rouven.essig@stonybrook.edu}
\affiliation[a]{C.N. Yang Institute for Theoretical Physics, Stony Brook University, Stony Brook, NY 11794, USA}

\author[b]{Jeremy Mardon,}
\emailAdd{jmardon@stanford.edu}
\affiliation[b]{Stanford Institute for Theoretical Physics, Department of Physics, Stanford University, Stanford, CA 94305, USA}

\author[c]{Oren Slone,}
\emailAdd{shtangas@gmail.com}
\affiliation[c]{Raymond and Beverly Sackler School of Physics and Astronomy, Tel-Aviv University, Tel-Aviv 69978, Israel}

\author[c]{Tomer Volansky}
\emailAdd{tomerv@post.tau.ac.il}

\preprint{YITP-SB-16-27}

\abstract{
We explore a new low-threshold direct-detection concept for dark matter, based on the breaking of chemical bonds between atoms. 
This includes the dissociation of molecules and the creation of defects in a lattice.
With thresholds of a few to 10's of eV, such an experiment could probe the nuclear couplings of dark matter particles as light as a few 
MeV. 
We calculate the expected rates for dark matter to break apart diatomic molecules, which we take as a case study for more general systems.  
We briefly mention ideas for how chemical-bond breaking might be detected in practice.
We also discuss the possibility of detecting solar neutrinos, including pp neutrinos, with this experimental concept.  
With an event rate of $\mathcal{O}$(0.1/kg-year), large exposures are required, but measuring low-energy solar neutrinos would 
provide a crucial test of the solar model.  
}

\begin{document}

\maketitle
\flushbottom

\newpage

%%%%%%%%%%%%%%%%%%%%%%%%%%%
\section{Introduction} 
\label{sec:introduction}
%%%%%%%%%%%%%%%%%%%%%%%%%%%

In the last several decades, a wide range of evidence has been accumulating for the existence of dark matter (DM). 
So far, DM has manifested itself only via its gravitational interactions with ordinary matter, and its precise nature remains unknown. 
The most studied DM paradigm is the Weakly Interacting Massive Particle (WIMP). 
An interesting feature of the WIMP is the ``WIMP miracle'', namely that the thermal history of the WIMP predicts the correct 
relic abundance, provided that its mass and couplings are similar to those of the Standard Model $W$ and $Z$-bosons. 
This makes the WIMP compelling, since theoretical arguments related to the Higgs hierarchy problem independently suggest
new physics near the Weak scale, which is often accompanied by a stable DM candidate, the WIMP.  Moreover, 
WIMPs are readily discovered with current technologies.
In recent years, there has been an extensive effort to search for WIMPs directly with underground detectors, 
satellites and earth-based telescopes, and at colliders such as the LHC. 
These searches have not found any unambiguous evidence for DM.  This, together with the null results for 
new-physics searches at the LHC thus far, and the fact that there are many possible DM candidates besides the WIMP, urges us to  
significantly broaden our search for DM. 

Many alternative, non-WIMP theories have been studied in recent years, often pointing to light DM (LDM) 
below the GeV scale~\cite{Boehm:2003hm, Boehm:2003ha, Pospelov:2007mp, Feng:2008ya, Hooper:2008im, Morrissey:2009ur, Kusenko:2009up, Kaplan:2009ag, Essig:2010ye, Essig:2011nj, Choi:2011yf, Falkowski:2011xh, Lin:2011gj, Hochberg:2014dra, Hochberg:2014kqa, Boddy:2014yra,Rajagopal:1990yx, Covi:1999ty}. 
In this paper, we focus on the direct detection of DM using chemical-bond breaking between atoms in a molecule or a crystal, as first 
suggested in~\cite{Essig:2011nj}.  This could probe {\it DM-nucleon interactions} for DM as light as $\sim\!10-100$~MeV.  
Only one analysis,  by the CRESST collaboration, has probed below the GeV scale, to 500~MeV~\cite{Angloher:2015ewa}, although the planned the SuperCDMS SNOLAB experiment aims to probe to about 300~MeV~\cite{Agnese:2015ywx}.   
Previous direct detection studies that place bounds on LDM, e.g.~\cite{Essig:2012yx}, have considered DM-electron scattering, which 
allows one to probe down to the MeV scale~\cite{Essig:2011nj,Essig:2012yx,Essig:2015cda,Graham:2012su,Lee:2015qva} and perhaps even lower~\cite{Hochberg:2015pha, Hochberg:2015fth, Hochberg:2016ajh, Hochberg:2016ntt}, but assume that DM couples to electrons.
New detection methods involving chemical-bond breaking could significantly improve upon current detection capabilities, may allow for the detection of cosmic and solar axions, and may ultimately also be able to probe the lowest energy part of the solar neutrino spectrum.   For other recent proposals see~\cite{Va'vra:2014tia, Va'vra:2016srr, Schutz:2016tid, Kouvaris:2016afs}.

The typical binding energy of atoms is a few to 10's of eV, so that the scattering of sub-GeV DM with a nucleus can break the 
chemical bond in a molecule or crystal.  
The rate of such scatterings depends on the internal structure of the target, which can be captured in a non-trivial 
target-dependent form factor. 
In what follows, we discuss the physics of such scattering events and calculate the expected dissociation rates for DM scattering off a diatomic molecule. 
These are a proxy for various experimental setups, including the bond breaking of a nucleus within a crystal target, which may provide for a more realistic setup.  

We further discuss the scattering of solar neutrinos off the same targets. Solar neutrinos will eventually become an irreducible background for any DM direct detection experiment, but are an interesting signal 
by themselves.  
Indeed, the solar neutrino fluxes have only been measured partially~\cite{Abdurashitov:2009tn, Bellini:2013lnn, Gando:2014wjd, Aharmim:2011vm}, in particular the pp-neutrino component, which dominates the low-energy neutrinos. 
Moreover, the neutral current interactions of the low-energy spectrum have never been fully measured. 
Their detection is of great interest and would constitute a crucial test of our understanding of the solar model~\cite{Bahcall:2000nu} and the MSW effect\footnote{Note that as opposed to the well-measured $^7$Be and $^8$B neutrinos, the pp-neutrinos admit a non-adiabatic propagation in the sun and their oscillation probabilities are therefore expected to be distinct from the former.}~\cite{Wolfenstein:1979ni}.

Our study is focused on calculating the theoretical rates and discussing the properties of the DM and neutrino signals.  
In this paper, we do not present a proposal for an actual experiment, but our chosen diatomic molecules capture a range of 
possible targets and thus serve as useful examples.  
Moreover, we do discuss aspects of how one might be able to see a signal from a DM- or neutrino-induced chemical bond breaking.  
Any particular idea will have to deal with environmental and detector backgrounds, and it is currently not clear which one is most realizable.  
Crystals are of particular interest, since defects induced within a crystal by a DM scattering event can alter its spectroscopic properties.   Spectroscopic measurements of crystals may thus allow for real-time detection of DM interactions.  
Ideally, the nature of the scattering process in a particular target may allow for background versus signal discrimination on an 
event-by-event basis, much as in existing direct detection experiments, but a thorough discussion of this is beyond the scope 
of the paper and is left for upcoming publications.

The paper is organized as follows.  In Sec.~\ref{sec:principles}, we discuss the general principles required for the detection of light DM and solar neutrinos through chemical bond breaking. The formalism is presented in detail in Sec.~\ref{sec:Scattering}. We discuss the physics that is involved in the calculations and consider a number of approximations that allow for an easy understanding of the expected rates. In Sec.~\ref{sec:LightDM}, we present expected rates and potential sensitivities 
on DM detection. In Sec.~\ref{sec:Neutrinos}, we present expected rates for solar neutrino detection and calculate the neutrino floor for this class of experiments. We summarize in Sec.~\ref{sec:setups} and discuss how realistic experimental setups could be achieved.

%%%%%%%%%%%%%%%%%%%%%%%%%%%
\section{Principles of Detection} 
\label{sec:principles}
%%%%%%%%%%%%%%%%%%%%%%%%%%%

Most nuclear-recoil DM searches have energy thresholds above $\sim 1$~keV, not allowing for sensitivity to DM below the GeV scale.  
Only DAMIC~\cite{Aguilar-Arevalo:2015rum}, CDMSlite~\cite{Agnese:2015nto}, and CRESST-II~\cite{Angloher:2015ewa} have achieved 
thresholds of $0.3-0.5$~keV, allowing CRESST-II to achieve sensitivity down to DM masses of $m_{\chi} \simeq 500$~MeV. 
The sensitivity is reduced since for elastic $2\to 2$ LDM-nucleus scattering, the recoil energy, $E_{\rm elastic}$, of a nucleus with mass $m_{\rm nuc}$ is 
\bea\label{eq:ER}
E_{\rm elastic} \le \frac{2 \mu^2_{{\chi},{\rm nuc}} v^2}{m_{\rm nuc}} \simeq 5~{\rm eV}\, \left(\frac{m_{\chi}}{100~{\rm MeV}}\right)^2\, \left(\frac{28~{\rm GeV}}{m_{\rm nuc}}\right)\,,
\eea
which is suppressed by $m_{\rm nuc}$ and decreases quadratically with $m_{\chi}$.  
In Eq.~(\ref{eq:ER}), the second equality assumes that the nuclear mass is roughly that of silicon, which is used by several experimental groups. 
We also used the {\it maximum} DM-nucleus relative velocity, which is given by $v_{\rm esc}+v_{\rm Earth}$, where we take the DM escape velocity from the galaxy to be $v_{\rm esc} = 600$~km/s and the mean Earth velocity around the galactic center to be $v_{\rm Earth} = 240$~km/s \cite{Lewin:1995rx}. 
The energy, $E_{\rm elastic}$, is to be contrasted with the total energy available for the scattering, which is given by the kinetic energy of the LDM 
particle. 
The maximum kinetic energy is $\mdm v^2 / 2 \simeq 0.37 \units{keV}$~($\mdm$/100~MeV),
which is significantly larger than $E_{\rm elastic}$ for the same DM mass.  
Moreover, note that the nuclear recoil energy is not directly detectable. 
Depending on the material, it is usually converted to some combination of phonons, ionization, and scintillation, which are detected. For elastic LDM-nucleus recoils, this conversion is inefficient and typically only a few 10's of percent of 
$E_{\rm elastic}$ is directly measurable in current experiments.  

This brief discussion suggests that detecting LDM with masses much below $\sim 1$~GeV with current experiments 
searching for {\it elastic nuclear} recoils is very challenging.
Instead, {\it inelastic} processes can greatly enhance the sensitivity to low DM masses.   
In~\cite{Essig:2011nj}, several processes were suggested.  For example, by scattering off bound electrons, which typically have much 
higher speeds than the DM, all the LDM kinetic energy is in principle accessible.  
This leads to detectable signals in current noble-liquid target experiments~\cite{Essig:2011nj,Essig:2012yx}.  
Technology that is under development and may be available in the near future could also allow for the detection of single- or few-electron 
events in semiconducting targets~\cite{Essig:2011nj,Essig:2015cda, Graham:2012su, Lee:2015qva}, and single- or few-photon  
events in scintillating targets~\cite{Essig:2011nj,Derenzo:2016fse}.  
Future possibilities include the use of superconducting and two-dimensional 
targets~\cite{Hochberg:2015pha,Hochberg:2015fth,Hochberg:2016ntt}. 
Searching for DM-induced nuclear recoils can be done with conventional detectors~\cite{Kouvaris:2016afs} 
and is under development using superfluid helium~\cite{Carter:2016wid,Schutz:2016tid}.  

Another possibility, suggested in~\cite{Essig:2011nj}, is that inelastic DM-nucleus scattering may dissociate molecules. Here we focus on such chemical-bond breaking interactions, examining them in more detail. A potential handle to detect such interactions is that they could trigger a physical or chemical change, which could be much simpler to observe than a slow-moving recoiling nucleus. The observed signal depends on the target material and the precise experimental setup. 
However, the scattering rate that triggers bond breaking is expected to be rather independent of the details of the binding potential. Indeed, we will show that it depends sensitively on the binding potential's depth, which is determined by the binding energy. As a consequence, one may derive a general formalism to describe such processes, relevant for a wide range of experimental setups. 

Bond-breaking interactions exhibit typical dissociation energies thresholds 
of up to tens of eV and probe the DM-nucleon coupling. The basic ingredients required for the detection of sub-GeV LDM or other feebly-interacting particles are:
\begin{itemize}
\item \textbf{Low dissociation energy} for breaking of the chemical bond, ideally of order $\lesssim 10 \text{ eV}$. This energy is strongly dependent on the binding energy and masses of the target particles involved.
\item \textbf{A suitable target material}.  Two competing effects exist as the mass of the target material is raised: (i) enhancement of the reaction rate by coherence effects (for spin independent reactions), and (ii) suppression of the corresponding recoil energy. The choice of material may be influenced by the type of particle we wish to detect. For instance, heavy atoms may be more appropriate for detecting solar neutrinos, which are typically more energetic than e.g.~100 MeV LDM, while lighter atoms may be better suited for detecting LDM. 
\item {\bf An enhancement mechanism} of the signal, which can allow for the detection single or few bond-breaking event (see Sec.~\ref{sec:setups} for an example).
\item \textbf{Background discrimination} achieved by the ability to differentiate between low-energy (signal) and high-energy (background) events and between low-energy nuclear recoils (signal) and low-energy electron recoils (background).
\end{itemize}

Observing a single or a few bond-breaking events is challenging and requires an enhancement mechanism. 
 Although a thorough study of possible setups and corresponding backgrounds is beyond the scope of this paper, a discussion of few related aspects is given in Sec.~\ref{sec:setups}.  A more detailed study, which employs the detection of color-centers in crystalline detectors will be presented in~\cite{color-center, color-center-theory}.

Below, we develop the general formalism to calculate bond-breaking due to DM or neutrino scattering off nuclei.  For concreteness, we consider a system of  diatomic molecules, which have typical binding energies of $\DE \sim 1 -10 \text{ eV}$.  In particular, we choose 
hydrogen (H$_2$), nitrogen (N$_2$), and beryllium oxide (BeO), which are meant to be representative of the range of nuclear masses and binding energies of diatomic molecules ($\Delta E_B \simeq$ 4.75~eV, 9.79~eV, and 4.54~eV, respectively). We will also consider 
hypothetical versions of these molecules with the same nuclear masses but different binding energies to illustrate the dependence of the 
scattering rates and other quantities on these parameters.  We note that none of these elements may be ideal experimentally, but they 
illustrate the basic physics involved in molecular dissociation. Moreover, diatomic molecular dissociation as a means of LDM detection may itself  
not be ideal experimentally, but is representative of the physics involved in more general chemical-bond breaking 
interactions using more realistic target materials such as multi-atomic molecules and possibly dissociation of nuclei within crystals.

%%%%%%%%%%%%%%%%%%%%%%%%%%%
\section{Inelastic  Scattering}
\label{sec:Scattering}

We now present the formalism required for calculating the dissociation of a diatomic molecule due to scattering with a weakly interacting particle. Consider a molecule that consists of two nuclei with masses $m_1$ and $m_2$ bound by some potential. The weakly interacting particle scatters with nucleus $m_1$ transferring momentum $\qv$. If the energy transfer is large enough, the atoms escape the potential and dissociate, each with some final momentum $\mathbf{k_1}$ and  $\mathbf{k_2}$. The unbound molecule can be thought of as a single state with some center of mass (COM) energy, $E_r$, and momentum, $\qv = \mathbf{k_1} + \mathbf{k_2}$. One may also define an internal energy, $E_{\rm int}$, and momentum $\qtv = \frac{\mu_{12}}{m_1}\mathbf{k_1} - \frac{\mu_{12}}{m_2}\mathbf{k_2}$ (as measured in the molecule's COM reference frame) with $\mu_{12}$ the molecular reduced mass. The relations between these energies and momenta are,
\begin{eqnarray}
E_r & = & \frac{q^2}{2(m_1+m_2)} \nonumber \\
E_{\rm int} & = & \frac{\qt^2}{2 \mu_{12}}\,. \
\end{eqnarray}
Below, we present the physics involved in calculating the rate for such scattering events.

For the LDM masses we are considering, the de Broglie wavelength of the DM is typically larger than the nucleus and smaller than the distance between the nuclei within the molecule. Thus, the DM always interacts with a \textit{specific} nucleus within the molecule. For a given total target mass, the event rate for scattering with a certain element $m_1$ is proportional to the number of nuclei of type $m_1$. For diatomic molecules with identical nuclei, the recoil spectrum and event rate are identical for both nuclei and the total rate is enhanced by a factor of 2. For diatomic molecules with distinct nuclei, both the cross section and the kinematics differ for the two cases where the DM interacts with either nucleus. In this case, there are distinct spectra and event rates for each kind of interaction, which may or may not be distinguishable depending on the specific details of the experimental setup. In this study, we present event rates and spectra for one kind of interaction per molecule. Specifically, for H$_2$ and N$_2$ the factor of 2 is present in all results, whereas for BeO all results assume interactions with Be only. This gives intuition as to what is expected for any diatomic molecule.

We first consider a classical calculation, which is typically sufficient, 
since nuclei are heavy particles bound in a weak potential. 
We then consider quantum effects, which are important near or below the  classical dissociation thresholds. 
Quantum effects cause a small lowering of the minimum DM mass needed for dissociation, typically by $\mathcal O(10\%)$. 
We discuss both the full quantum calculation (for isolated diatomic molecules), and approximations, which allow for a simpler computation and an intuitive understanding of the effects. 
This intuition, and some of the formalism, should extend to more general targets. 
Explicit formulae for the total rates expected from DM and neutrino interactions will be presented in Secs.~\ref{sec:LightDM} and~\ref{sec:Neutrinos}, respectively.

\subsection{The Classical Limit}
The scattering of an LDM particle, $\chi$, with a diatomic molecule, dissociating the molecule into its constituents, can be treated as an inelastic $2 \rightarrow 2$ process. In this picture, the initial state consists of the incoming DM particle and the bound molecule, while the final state consists of the outgoing DM particle and the unbound molecule. If the final internal energy is much larger than the binding energy of the system, the angular momentum of the final state is a quantum number much larger than unity, and the system may be regarded classically. If the internal energy is of order, or smaller than, the binding energy of the system, then the importance of quantum effects is determined by the interplay between the momentum spread of the initial wavefunction of the molecule, the binding energy, and the reduced mass of the molecule (details below). At these low internal energies, quantum effects become important for large values of the momentum spread and small values of the binding energy and reduced mass.  

Classically, we can treat a diatomic molecule as two point particles sitting at the minimum of 
an attractive potential (which binds the two atoms). Energy conservation implies,
\beq
\qt^2 = \left( \frac{\mu_{12}}{m_1} \right)^2 q^2 - 2 \mu_{12} \DE\,,
\label{eq:Clas_Energy_Conservation}
\eeq
and dissociation occurs if the internal energy 
is larger than the binding energy, as long as there is no bound solution with non-zero angular momentum.  This translates to the following minimum momentum transfer from $\chi$ to the struck atom in the molecule,
\beq
q_\text{min} = \frac{m_1}{\mu_{12}} \sqrt{2\mu_{12} \DE}\,,
\label{eq:Class_Kinematics_1}
\eeq
where $\DE$ is the molecular binding energy. 
The kinematically allowed range of $q$ is,
\beq
0 < q \leq 2 \mu_{\chi 1} v\,,
\label{eq:Class_Kinematics_2}
\eeq
where $\mu_{\chi 1}$ is the reduced mass of the DM-$m_1$ system.  Eqs.~\eqref{eq:Class_Kinematics_1} and \eqref{eq:Class_Kinematics_2} are the kinematical constraints on the interaction rate.  The average differential cross section is,
\beq
\left<\frac{d\sigma v}{d q^2} \right> = \frac{\sigmaa}{4 \mu_{\chi 1}^2 v}\,,
\label{eq:Class dsigma_dq2}
\eeq
where $\sigmaa$ is the DM-nucleus interaction cross section with the following relation to the DM-nucleon interaction cross section,
$\sigmap$,
\beq
\sigmaa = \left[ f_PZ+f_N(A-Z) \right]^2 \frac{\mu_{\chi 1}^2}{\mu_{\chi n}^2} \sigmap\,.
\eeq
Here $A$ and $Z$ are the mass and atomic numbers of $m_1$, $f_P$ ($f_N$) is the coupling strength to the proton (neutron) and $\mu_{\chi n}$ is the reduced-mass of the DM-nucleon system. Throughout this study we take the nuclear form factor (usually taken to be the Helm form factor~\cite{Lewin:1995rx}) to be unity. This is a good approximation since the momentum transfer involved is much smaller than the typical binding energy of the nucleus. The $\left[ f_PZ+f_N(A-Z) \right]^2$ enhancement occurs because of a coherence effect, as long as the de Broglie wavelength of the incoming DM particle is larger than the typical size 
of the nucleus, as is typically the case for the systems we are considering.

\subsection{Quantum Effects}
\label{sec:quantum-effects}
The classical approximation often breaks down close to threshold when the energy transfer is of order the binding energy of the system. Understanding an experiment's sensitivity to the lightest possible DM masses thus requires gaining control over the quantum corrections. Quantum mechanically, the momenta of the individual atoms in the molecule are not definite. This allows for the atoms in the initial state to have a nonzero momentum, in contrast to the classical limit, so bond breaking can occur with less momentum transfer to the target than in the classical case. Consequently, Eq.~(\ref{eq:Class_Kinematics_1}) does not hold.  
In particular, even at threshold there is a spread in values of $q$ that allow for the dissociation of the target.

Quantum mechanically, the initial state has some typical momentum spread, $\Delta p$, which decreases the momentum transfer necessary for dissociation,
\beq
q^\text{quant}_\text{min} \simeq \frac{m_1}{\mu_{12}}\left( \sqrt{2\mu_{12}\DE} - \Delta p \right)\,,
\label{eq:Quant_qmin}
\eeq
whereas the minimal classical value, $q^\text{class}_\text{min}$, is given by Eq.~(\ref{eq:Class_Kinematics_1}). Thus, the ratio of the minimal quantum vs classical DM mass is,
\begin{eqnarray}
\frac{m^\text{quant}_\text{min}}{m^\text{class}_\text{min}} = \frac{q^\text{quant}_\text{min}}{q^\text{class}_\text{min}} & \approx & 1 - \frac{\Delta p}{\sqrt{2 \mu_{12} \DE}} \nonumber \\
& \approx & 1 - 0.2 \left(\frac{\Delta p}{4 \alpha m_e}\right) \left(\frac{0.5 \text{ GeV}}{\mu_{12}}\right)^{1/2} \left(\frac{5 \text{ eV}}{\DE}\right)^{1/2}\,,
\label{eq:m_Threshold}
\end{eqnarray}
where $m^\text{class}_\text{min}$, $m^\text{quant}_\text{min}$ are the classical and quantum mass thresholds, respectively. In the second line of Eq.~(\ref{eq:m_Threshold}) we have taken typical values for an H$_2$ molecule. A more localized initial wavefunction corresponds to larger values of $\Delta p$. The momentum spread is typically of order $\Delta p \sim \alpha m_e$ ($\alpha m_e$ is the inverse of the Bohr radius) since the typical uncertainty in distance between nuclei is $\Delta r_0 \sim (\alpha m_e)^{-1}$.

When quantum corrections become important, the classical calculation for the cross section is no longer valid and the quantum mechanical matrix element must be evaluated. Naively, scattering of DM or a neutrino with a 
molecular target is a $2 \rightarrow 3$ process. For instance, in the case of a diatomic molecule, the final states correspond to the outgoing DM or neutrino particle and the two dissociated atoms. The formalism that follows is relevant for any process of this type, i.e.~a process where the target is initially in a bound state and consists of free particles after the interaction occurs. 
However, much like the classical case discussed above, such a process can be thought of as a $2 \rightarrow 2$ scattering process with a non-trivial form factor. The matrix element for such a scattering can be written as,
\beq
\mathcal{M}_{fi}(\mathbf{q},\tilde{\mathbf{q}}) \equiv \mathcal{M}_{free}(\mathbf{q}) F\left(\frac{\mathrm{\mu_{12}}}{\mathrm{m_{1}}}\mathbf{q},\tilde{\mathbf{q}}\right),
\label{eq:M_Matrix_FF}
\eeq
where $\mathcal{M}_{free}$ is the matrix element for the elastic scattering of the incoming particle with a {\it free} nucleon and $F(\frac{\mathrm{\mu_{12}}}{\mathrm{m_{1}}}\mathbf{q},\tilde{\mathbf{q}})$ is the form factor which encodes the quantum information of the initial and final states of the target. 
The information regarding the binding potential of the target, and thus the true $2 \rightarrow 3$ process of interest, is parameterized by the form factor. That said, we show below that quantum information is not important in most energy-transfer regimes, 
and the process can be well approximated by a simple classical scattering. 

Assuming a weakly interacting particle scatters against a nucleon within the target, 
the form factor is given by,
\beq
F(\mathbf{q},\tilde{\mathbf{q}}) \equiv \int d^{3}re^{i\mathbf{\frac{\mathrm{\mu_{12}}}{\mathrm{m_{1}}}\mathbf{q}\cdot r}}\Psi_{\qtv}^{*}(\mathbf{r})\Psi_{i}(\mathbf{r})\,.
\label{eq:Fmol}
\eeq
The functions $\Psi_{\qtv}(\mathbf{r})$, $\Psi_{i}(\mathbf{r})$ are the COM final and initial molecular wavefunctions respectively.
Integrating over the final momentum defines the dimensionless {\it dissociation form factor} as
\beq
\Fdis = \frac{\qt^{3}}{(2\pi)^{3}} \int d\Omega_{\qt} |F(\frac{\mu_{12}}{m_{1}}\mathbf{q},\qtv)|^{2}\,.
\label{eq:Fdis}
\eeq
The evaluation of Eq.~(\ref{eq:Fdis})  involves solving the Schr\"odinger equation for the initial and final states of the target molecule, which in turn requires detailed knowledge of the binding potential. Fortunately, as we now argue and demonstrate in the next subsection, the form factor depends only mildly on the precise form of the potential, and the corresponding theoretical uncertainties regarding the specific shape of the potential for a given target are therefore low.

The minima of the binding potential are often well approximated by a harmonic potential, so that the radial wavefunction of the ground state of the target is well modeled by a harmonic solution. The full initial wavefunction takes the form 
\beq
\Psi_{i}(\mathbf{r}) = \mathcal{Y}_{\ell m}(\Omega) \left(\frac{1}{\sigma_{0}\sqrt{\pi}}\right)^{\frac{1}{2}} \frac{e^{-\frac{1}{2}\left(\frac{r-r_{0}}{\sigma_{0}}\right)^{2}}}{r}\,,
\label{eq:Psi_i}
\eeq
where $\mathcal{Y}_{\ell m}(\Omega)$ are spherical harmonics, $r_0$ is the distance between nuclei for which the potential finds its minimum and $\sigma_0$ is the variance of the wavefunction. The momentum spread, $\Delta p$, described above is proportional to $\sigma_0^{-1}$. As long as we are solving for the ground state, Eq.~(\ref{eq:Psi_i}) is a good approximation.
The form of the final wavefunction depends on the exact form of the binding potential. However, since the form factor calculation involves only the overlap of the initial and final state wavefunctions, only the form of the final wavefunction in the vicinity of the origin is of importance. 
This fact greatly simplifies the calculations for non-trivial binding potentials.  

\subsection{Spherical Symmetry}
\label{sec:rates-sph-symm}

For a spherically symmetric potential, the final wavefunctions can be expanded as 
\beq
\Psi_{\qtv}(\rv) = 4 \pi \sum_{\ell,m} \frac{a_{\ell}}{2\ell+1} \Ylm_{\ell m}^*(\Omega_{\qt}) \Ylm_{\ell m}(\Omega_r) R_{\qt \ell}(r)\,,
\label{eq:Psi_Expand}
\eeq
where $R_{\qt \ell}(r)$ is the radial solution for the final state and $\Omega_r, \Omega_{\qt}$ are the solid angles with respect to the coordinates $r$ and $\qt$ respectively.  Summing over all final angular momentum values, the form factor takes the form (see Appendix~\ref{Appendix:Full} for details),
\beq
\Fdisq = 8 \qt^3 \sum_{\ell} \frac{|a_\ell|^2}{(2\ell+1)} \left| \int dr r^2 j_{\ell}(\mum qr) R_{\qt \ell}(r) \Psi_i(r) \right|^2\,,
\label{eq:Full_FF_1}
\eeq
which must be solved in order to calculate the interaction rate.

Evaluating Eq.~(\ref{eq:Full_FF_1}) is non-trivial as it requires knowledge of the outgoing states for all values of the internal energy and angular momenta.  However, if we neglect the binding potential and use the Born approximation for the outgoing wavefunction, i.e.~take the outgoing state to be a plane wave, Eq.~(\ref{eq:Full_FF_1}) simplifies considerably. Corrections from the binding potential deform the outgoing wavefunction around the origin.  However, if the internal energy of the final state is much larger than the binding energy,  $E_{\rm int} \gg \DE$, the Born approximation is sufficiently accurate as is shown below.

The quantum mechanical calculation (unlike the classical calculation) takes into account the non-zero momentum of the initial state, which allows dissociation of a molecule with momentum transfer lower than the classical threshold given in Eq.~(\ref{eq:Class_Kinematics_1}). This enhances the rate near threshold with respect to the classical result. In the Born regime, assuming zero binding energy, the dissociation form factor, Eq.~(\ref{eq:Fdis}), takes the following analytical form 
(see Appendix \ref{Appendix:Born} for details),
\beq
\Fdisq = \frac{m_1}{\mu_{12}} \frac{\qt^2}{\pi^{3/2} \sigma_{0} q} \int_{\qt-\mum q}^{\qt+\mum q} \frac{dK}{K} \label{eq:Fdis_Born} \cdot \left| \int dr \sin(Kr) e^{-\frac{1}{2}\left(\frac{r-r_{0}}{\sigma_{0}}\right)^{2}} \right|^2\,.
\eeq
The function $\Fdisq$ has non-trivial support around the region
\beq
q \simeq \frac{m_1}{\mu_{12}}\qt = \frac{m_1}{\mu_{12}} \sqrt{2 \mu_{12} E_{\rm int}}\,,
\label{eq:q_qt_Born}
\eeq
which is just the classical result of energy conservation, Eq.~(\ref{eq:Clas_Energy_Conservation}), assuming zero binding energy.

To accurately calculate the rate for low DM masses, taking into account  the binding energy, an improvement on 
the Born approximation is required. 
This can be achieved simply by accounting for the binding energy in the internal momentum of the final state, $\qt$.  
This is equivalent to changing the limits of the integral in Eq.~\eqref{eq:Fdis_Born}: $\qt \to \qt^\prime = \sqrt{2 \mu_{12} (E_{int} + \DE)}$.
The form factor now has sizeable support in the region,
\beq
q \simeq \frac{m_1}{\mu_{12}}\qt^\prime = \frac{m_1}{\mu_{12}} \sqrt{2 \mu_{12} (E_{int} + \DE)}\,,
\label{eq:q_qt_Born_Fixed}
\eeq
which is the classical result for dissociation of a molecule with non-zero binding energy, Eq.~(\ref{eq:Clas_Energy_Conservation}). We denote this correction the {\em Improved Born Approximation}.

\begin{figure}[t]
\centering
Molecular Dissociation Form Factor\\
\vspace{4mm}
\includegraphics[width=0.483\textwidth]{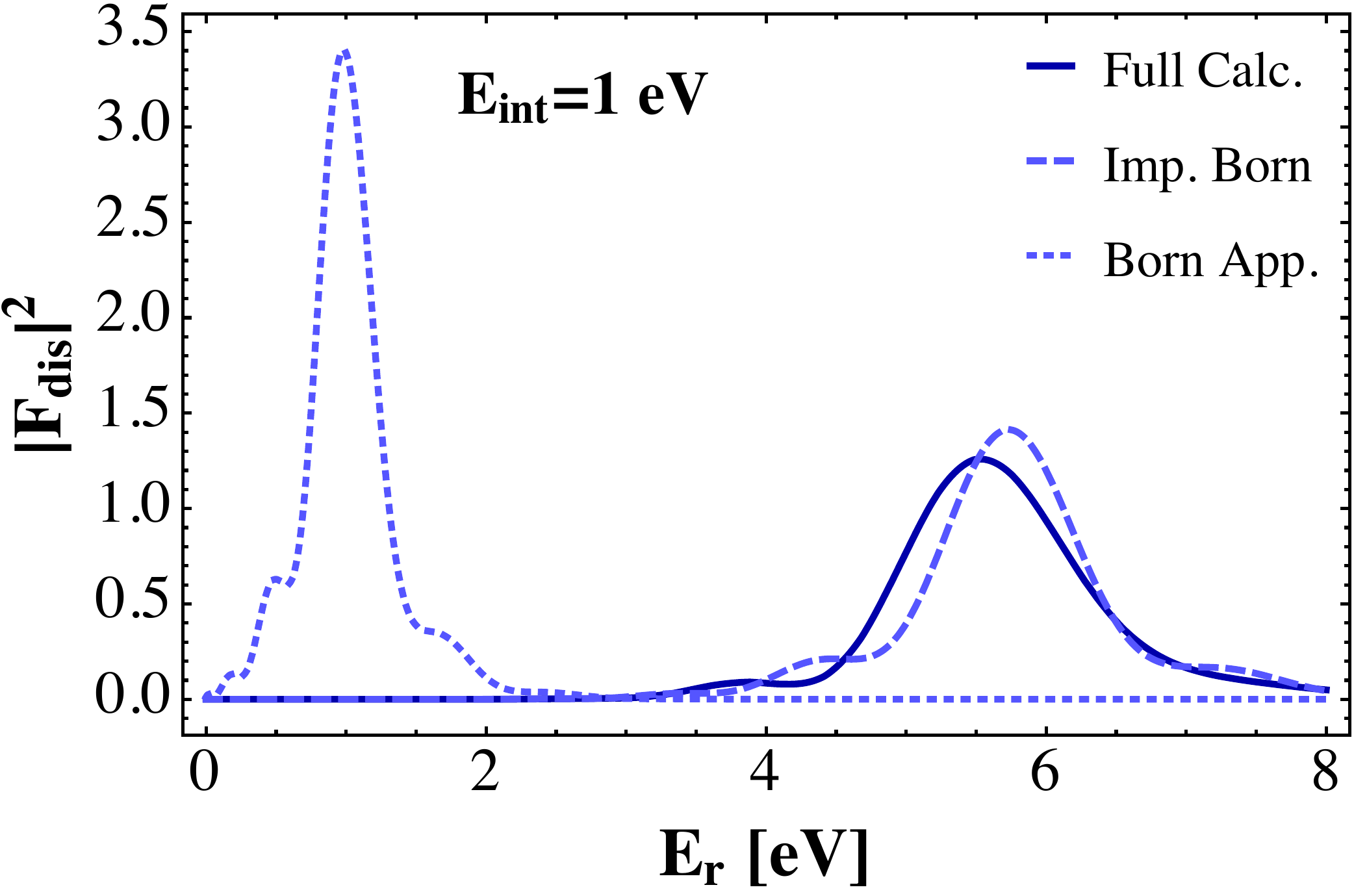}
\hspace{0.3cm}
\includegraphics[width=0.477\textwidth]{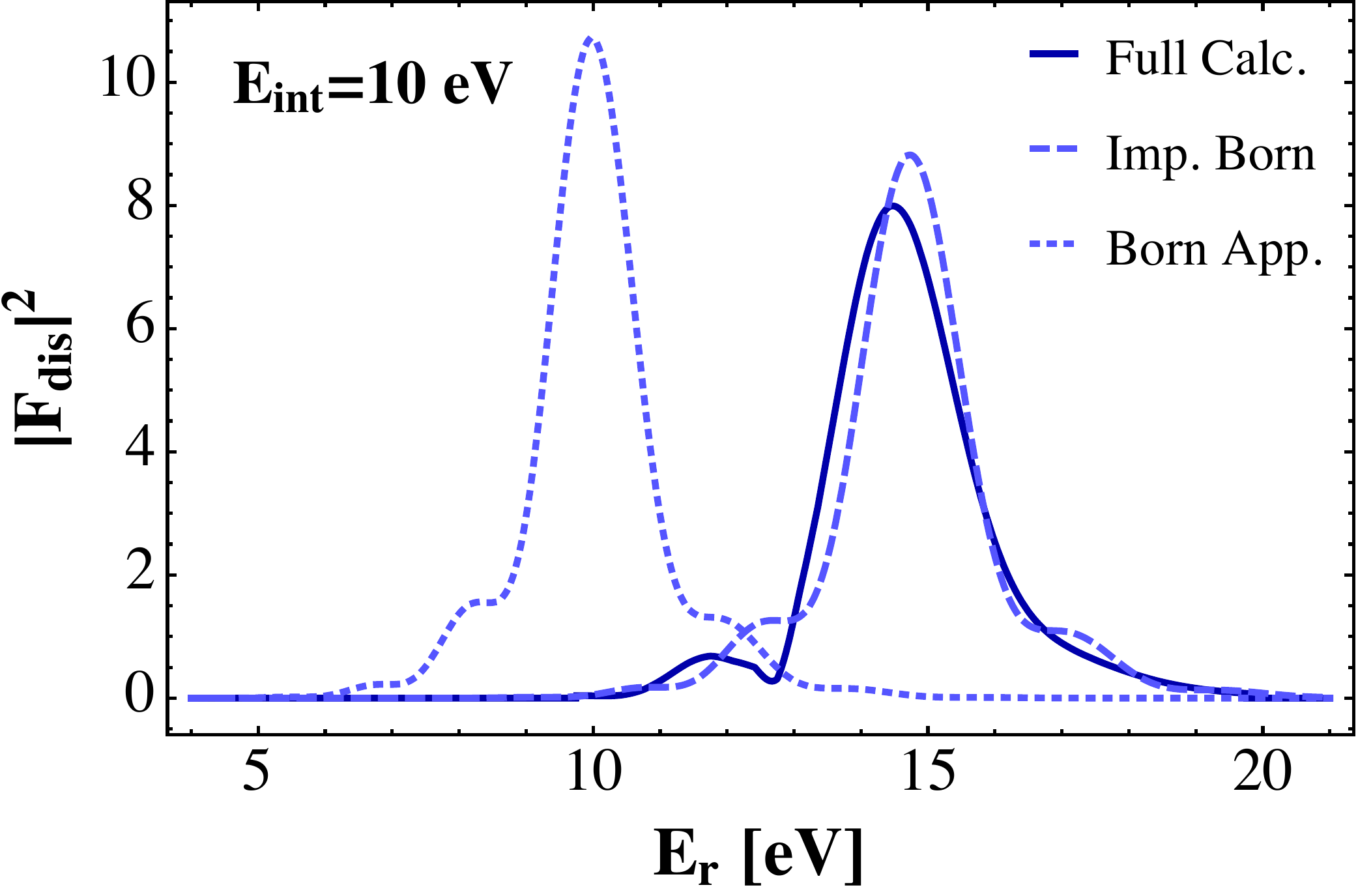}\\
\begin{minipage}[t]{0.5\textwidth}
\mbox{}\\[-0.1\baselineskip]
~~~~~~~\includegraphics[width=0.96\textwidth]{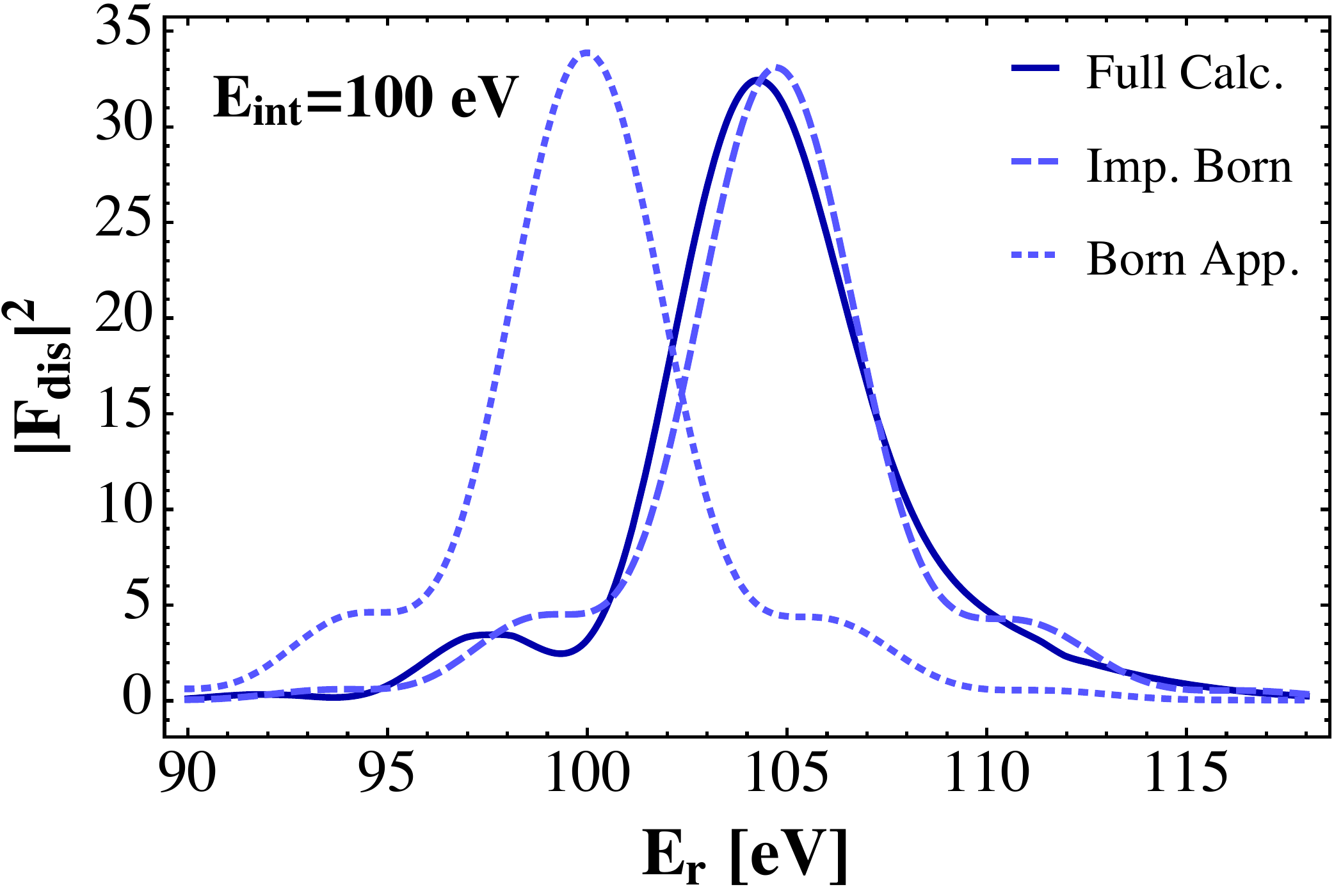} 
\end{minipage}
\hfill
\begin{minipage}[t]{\textwidth}
\mbox{}\\[-\baselineskip]
\caption{
Molecular dissociation form factors, $\Fdisq$, calculated in different approximations. 
The dotted and dashed curves correspond to the Born and Improved Born approximations respectively, while the solid curves correspond to the full quantum calculation using the full final state wave functions (see Sec.~\ref{sec:rates-sph-symm}). 
These results are for molecular hydrogen with binding energy $\DE=4.75\,\text{eV}$, with final-state internal energies of $1 \text{ eV}$ ({\it top left}), $10 \text{ eV}$ ({\it top right}), and $100 \text{ eV}$ ({\it bottom}).  The Improved Born approximation agrees well with the full calculation down to energies close to threshold, but is much easier to compute. 
\label{Fig:FF_Comparison}
}
\end{minipage}
\end{figure}

Accounting for the binding energy introduces two competing effects, one which enhances the rate and the other which acts to suppress it.
An enhancement occurs for nonzero $\Delta E_B$, since the minimum momentum required to dissociate the molecule is $\sqrt{2\mu_{12} (E_{\rm int} + \DE)}$ as opposed to $\sqrt{2\mu_{12} E_{\rm int}}$ in the Born approximation.  Consequently, the volume of available phase space is larger, which increases the dissociation rate.  This is directly related to the Sommerfeld enhancement (for a review see, e.g.~\cite{ArkaniHamed:2008qn}) occurring for an outgoing state~\cite{Essig:2011nj}.
A suppression of the rate occurs, since a larger momentum transfer is required to dissociate the molecule for nonzero $\Delta E_B$.  Larger momentum transfers require a larger initial DM velocity, allowing less DM particles in the halo to participate in the scattering.  
More precisely, the minimum DM velocity required for a scattering event to occur is 
\beq
v_{\rm min} = \frac{E_{\rm int}+\DE}{q} + \frac{q}{2 \mu_{\chi m}}\,,
\label{eq:vmin_General}
\eeq
where $\mu_{\chi m}$ is the reduced mass of the DM-target system (i.e.~the entire molecule for the case of DM-molecule scattering)\footnote{Note that $v_{\rm min}$ is minimized at $q_0 = \sqrt{2 \mu_{\chi m}(E_{\rm int} + \DE)}$.}. Put simply, if more energy is required for dissociation, a smaller region in the DM velocity profile can contribute to the dissociation.

A diatomic molecule is a suitable spherically symmetric case study with which one can understand the expected rates for various other target materials.  In the COM frame of a molecule, the potential is spherically symmetric and therefore Eq.~(\ref{eq:Psi_Expand}) holds.  Solving for the wavefunctions requires knowledge of the binding potential, $V(r)$, which can often be modeled by a Morse potential \cite{Morse:1929zz} with the form,
\beq
V_M(r) = \DE \cdot e^{-\alpha_0 (r-r_{0})} \left( e^{-\alpha_0 (r-r_{0})} -2 \right)\,,
\label{eq:V_Morse}
\eeq
where again $r_{0}$ is the minima of the potential and $\alpha_0 = \sqrt{\frac{V^{\prime\prime}(r_{0})}{2 \DE}}$.

Fig.~\ref{Fig:FF_Comparison} presents examples of the form factor calculated using either the Born or the Improved Born Approximation, for three values of $E_{\rm int}$ and for a Morse potential that approximates molecular hydrogen, H$_2$, with binding energy $\DE = 4.75 \text{ eV}$.  These have been compared with the full calculation for each energy, i.e.~Eq.~(\ref{eq:Full_FF_1}) with the exact solutions for the final wavefunctions. As the internal recoil energy increases, the three calculations converge to the same functional form.  
While the Born approximation does not hold for low values of $E_{\rm int}$, the Improved Born Approximation successfully mimicks the correct form factor for even low values of $E_{\rm int}$. It is therefore not necessary to fully calculate Eq.~(\ref{eq:Full_FF_1}). 

\begin{figure}[t]
\centering
\includegraphics[width=0.43\textwidth]{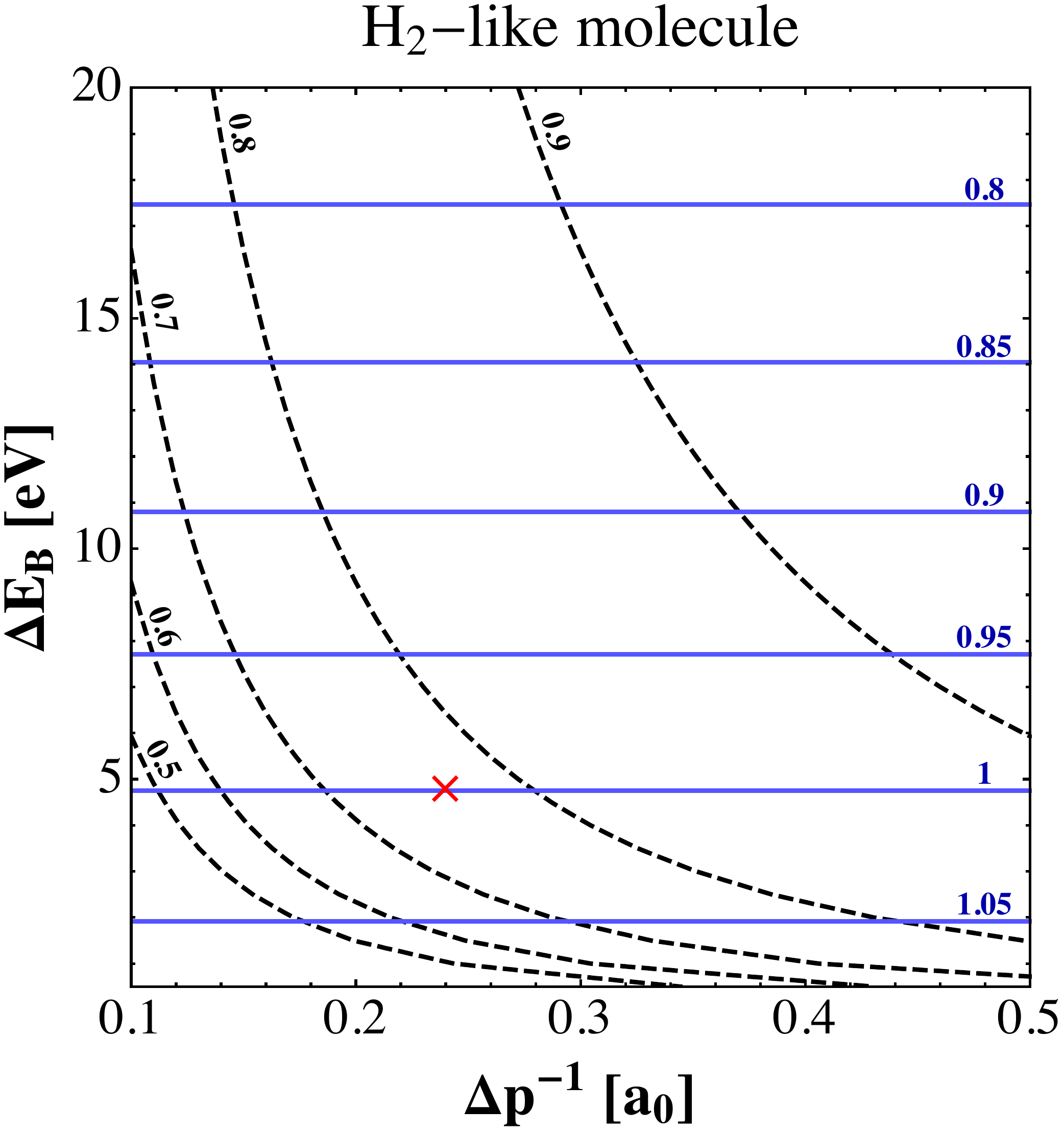}
\hspace{0.2cm}
\includegraphics[width=0.43\textwidth]{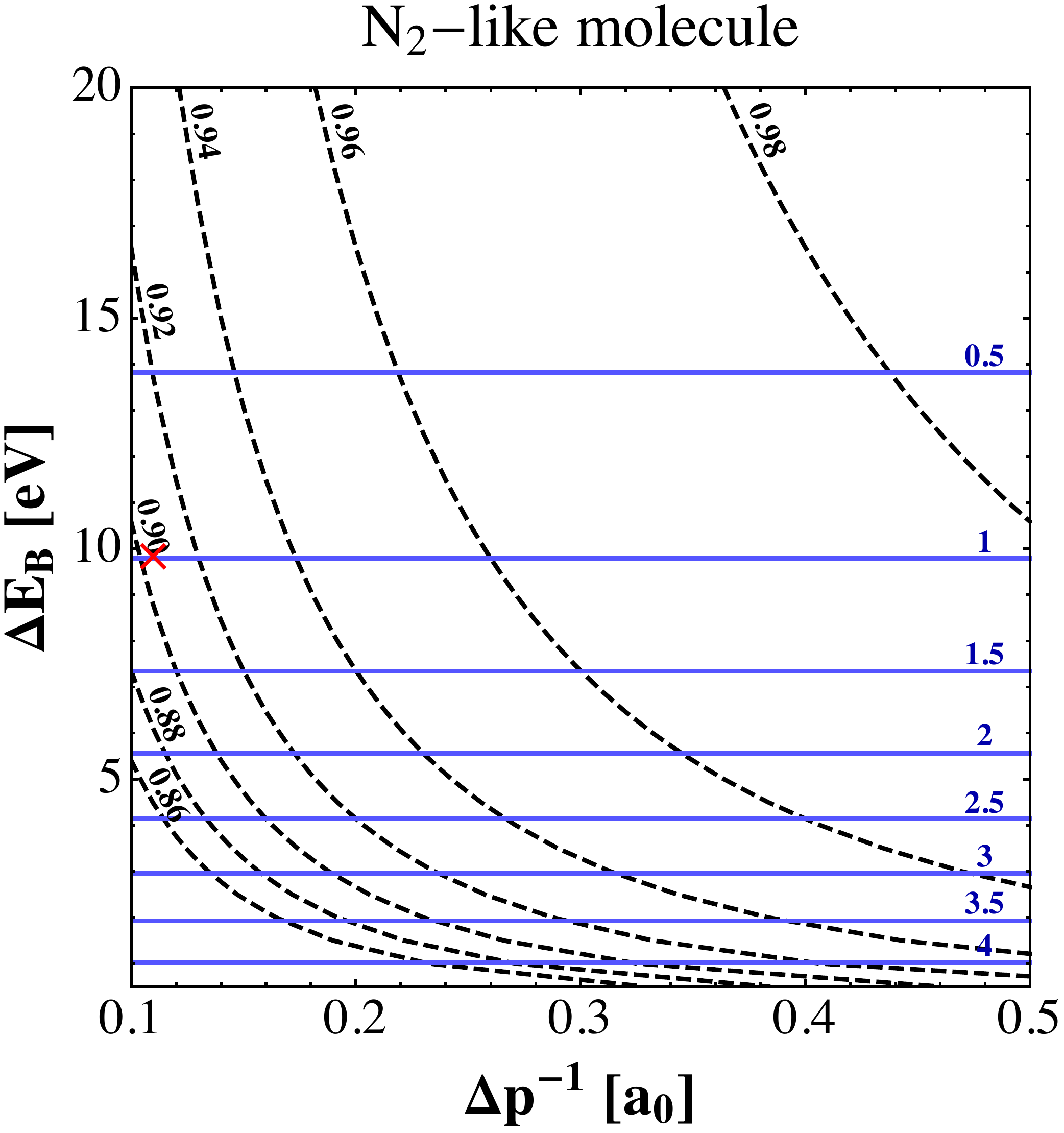}\\
\begin{minipage}[t]{0.5\textwidth}
\mbox{}\\[-0.01\baselineskip]
~~~~~~~\includegraphics[width=0.852\textwidth]{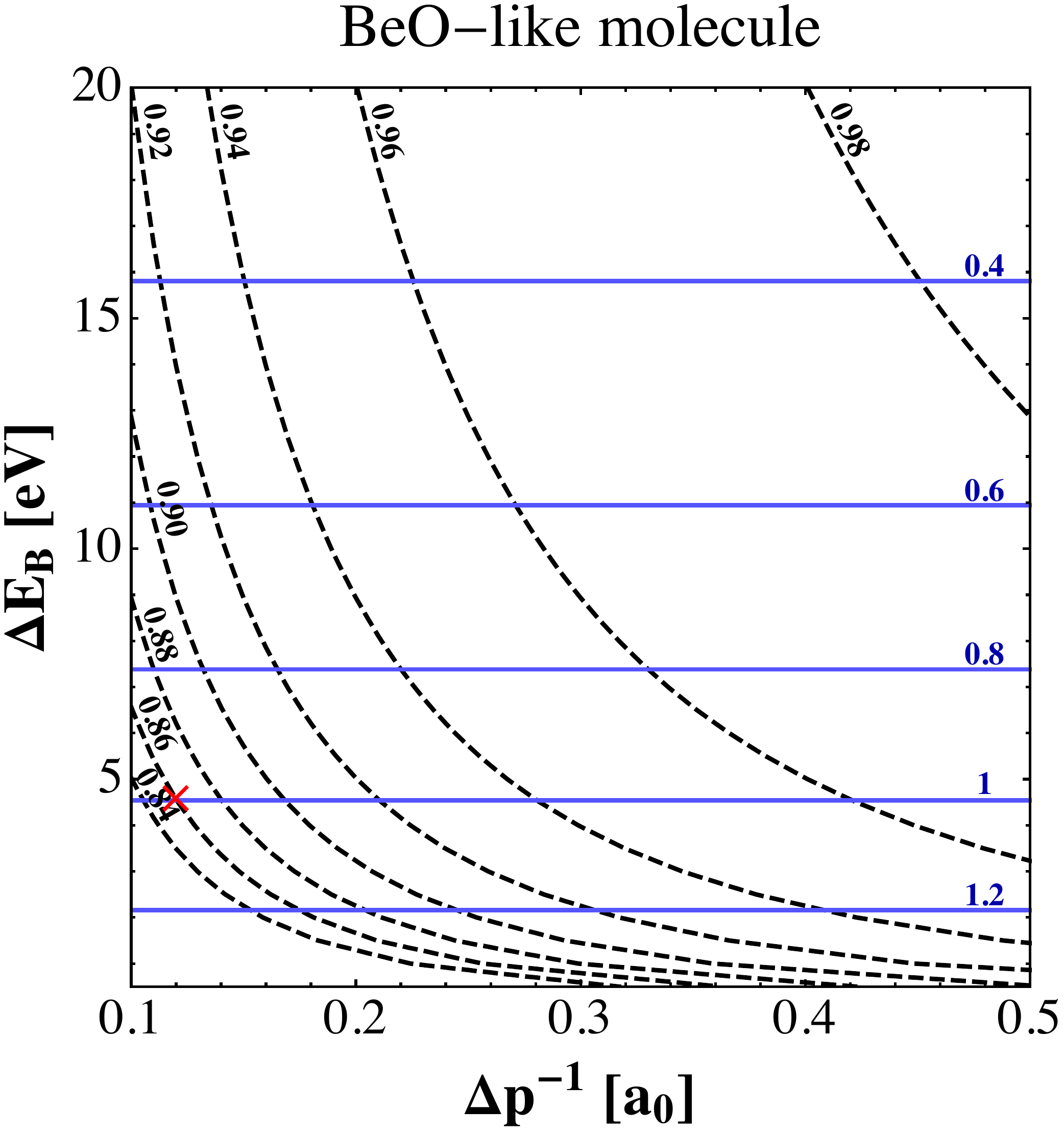} 
\end{minipage}
\hfill
\begin{minipage}[t]{\textwidth}
\mbox{}\\[-\baselineskip]
\caption{
Dependence of dissociation rates and quantum-to-classical mass thresholds on the momentum spread of the initial wave-function, $\Delta p$, and the molecular binding energy, $\DE$. 
{\it Blue contours} show the dissociation rates for a given diatomic molecule obtained when varying $\Delta p$ and $\DE$, 
normalized to the rates when $\Delta p$ and $\DE$ are set to their correct values found in Nature. 
A {\it red cross} on each panel marks the values found in Nature. 
{\it Dashed black contours} present the ratio of the quantum-to-classical mass thresholds, $m_{\text{min}}^{\text{quant}} / m_{\text{min}}^{\text{clas}}$, see Eq.~(\ref{eq:m_Threshold}). 
We show three diatomic molecules H$_2$ ({\it top left}), N$_2$ ({\it top right}), and BeO ({\it bottom}), and choose the DM mass $m_\chi = 300 \text{ MeV}$. 
We see that the dissociation rates are nearly independent of $\Delta p$ but depend sensitively on $\DE$, while  
quantum corrections are affected by both parameters, especially for lighter elements. 
\label{fig:Param_Dependence}
}
\end{minipage}
\end{figure}

We now show that the most important property of the binding potential that determines the dissociation rate is the binding energy, while 
the detailed shape is less relevant.  As mentioned above, this is because the form factor is largely determined by the overlap of the initial and final wavefunctions.  The initial wavefunction is localized around the minimum of the potential, and the shape of the binding potential does not have a sizable effect on the final wavefunction around the origin.  For a Morse potential, this is equivalent to accounting for the dependence of the solution on $\DE$ and neglecting the dependence on $r_{0}$ and $\alpha_0$.  In Fig.~\ref{fig:Param_Dependence}, we present the dissociation rate (\textit{blue}) for a given DM mass, $m_\chi=300 \text{ MeV}$ for our three representative diatomic molecules: H$_2$, N$_2$, and BeO. The results are shown as a function of $\DE$ and the inverse of the momentum uncertainty, $\Delta p^{-1}$. As discussed above, $\Delta p^{-1}$ is proportional to the variance, $\sigma_0$, of the initial wavefunctions, which we take to be gaussian functions. The figure shows the resulting dissociation rates normalized to the expected rate for the same molecule with its parameters set at their correct values (in each panel the correct values are marked by a red colored cross). The results show that there is only extremely mild dependence on the shape of the initial wavefunction and that the effect depends mostly on the binding energy. In the same figure, the \textit{dashed black} lines show the ratio of the quantum-to-classical mass threshold, Eq.~(\ref{eq:m_Threshold}), for the same parameter space.

\subsection{Non-spherical symmetry}
Bond breaking in a crystal lattice (as opposed to a diatomic molecule) poses a number of complications to the physics described above. One complication is related to the fact that the potential that binds the nucleus to the crystal bulk is non-spherically symmetric and therefore, in principle, Eq.~(\ref{eq:Psi_Expand}) no longer holds. However, as explained above, the dissociation rate is only mildly dependant on the form of the binding potential and depends almost solely on the binding energy of the system. This is expected to remain true for a non-spherical symmetric binding potential as long as the overlap of the initial and final wavefunctions is highly localised around the origin. In such a case, one needs only to obtain the form of the outgoing target wavefunction around the origin. 

A second modification occurs, since the kinematics and the binding energy of the system are no longer straightforward. 
Since the potential is non-spherical, the classical trajectory of a scattered particle  depends on its initial momentum,  scanning the potential in a directionally-dependent manner.
Writing the rates as a function of the binding energies is therefore inconvenient.  
Instead, it is simpler to discuss the (well-defined) minimal recoil energy in a given direction,  $E_r^{\rm min}(\Omega_q) = q_{\rm min}^2 (\Omega_q) / 2 M$, needed in order to dissociate the molecule.  
The directionally-dependent minimal energy follows from calculating the classical path for a dissociated nucleon initially at the potential minimum, after receiving a momentum kick. Each classical path takes into account the dissipative multi-scattering process and can be mapped to such a minimal recoil energy.  
Correspondingly, the RHS of $q_{\rm min}$ in Eqs.~\eqref{eq:Class_Kinematics_1} and \eqref{eq:Quant_qmin}, is replaced by $q_{\rm min}(\Omega_q)$.
 
With the knowledge of the function $E_r^{\rm min}(\Omega_q)$  and if the initial wavefunction is  localized and is approximately spherically symmetric, the dissociation rate may be reduced to the same form as that of the spherical symmetric case with the kinematics and binding energy modified accordingly, and a dependence on the direction of $\qv$. In principle, this should create a modulation in a DM or neutrino signal that depends on the orientation of the incoming particles with respect to the orientation of the target.

We conclude that calculating the rates for a crystal target requires taking into account the time dependence of the potential, 
modeling of the binding potential, and secondary effects after dissociation. This calculation is beyond the scope of this paper, but will be discussed in a future publication.

%%%%%%%%%%%%%%%%%%%%%%%%%%%
\section{Dissociation Rates for Light Dark Matter} 
\label{sec:LightDM}
%%%%%%%%%%%%%%%%%%%%%%%%%%%

\subsection{Rates and Potential Sensitivities}
In order to parametrize the cross section and scattering rate in a model-independent way, we define a DM form factor, $F_{\rm DM}(\mathbf q)$, and a reference cross section, $\sigmap$, as follows,
\begin{eqnarray}
\FDM & \equiv & \frac{|\mathcal{M}_{2\to2}(\mathbf{q})|^{2}}{|\mathcal{M}_{2\to2}(\mathbf{q}^{2} = q_0^2)|^{2}} \label{eq:DD_FF} \,, \\
\sigmap & \equiv & \frac{|\mathcal{M}_{2\to2}(\mathbf{q}^{2} = q_0^2)|^{2}}{16\pi(m_{\chi}+m_{n})^{2}}\,.
\label{eq:sigma_p}
\end{eqnarray}
Here $q_0$ is a fixed value of the momentum transfer, which we take to be $q_0=100 \text{ keV}$ throughout. 
$\mathcal{M}_{2\to2}(\mathbf q)$ is the matrix element for free scattering of DM with a nucleon with momentum transfer $\mathbf q$, and $m_n$ is the nucleon mass. All the dependence on the modeling of the DM sector is encapsulated in $\FDM$ and $\sigmap$. For a trivial DM form factor, i.e. $\FDM=1$, $\sigmap$ is just the free scattering cross section of DM with a nucleon.

As discussed above, the dissociation rate can be treated as a classical or a quantum process depending on the energy transfer to the target. In the classical regime, the dissociation rate is simply given by 
\beq
R = \left[ f_PZ+f_N(A-Z) \right]^2 N_T \frac{\rho_\chi}{m_\chi} \frac{\sigmap}{4 \mu_{\chi n}^2} \int dq^2 \times \FDM \Theta\left(q - \sqrt{\frac{2 m_1^2}{\mu_{12}} \DE}\right) \eta(v_{\rm min})\,,
\eeq
where $N_T$ is the number of target atoms, $\rho_\chi$ is the DM density, $m_\chi$ is the DM mass, and we have used the average differential cross section from Eq.~\eqref{eq:Class dsigma_dq2}. The function $\eta(v_{min})$ is 
\beq
\eta(v_{\rm min})=\int_{v_{\rm min}}^\infty d^3 v \frac{\fv}{v}\,,
\eeq
where $\fv$ is the velocity distribution profile of the DM in the Milky-Way halo (see e.g.~\cite{Lewin:1995rx}) with $v_0 = 230 \text{ km/s}$, $v_{\rm Earth} = 240 \text{ km/s}$, and $v_{\rm esc} = 600 \text{ km/s}$. The minimum velocity is Eq.~(\ref{eq:vmin_General}) with the classical value of $E_{\rm int}$, i.e. $v_{\rm min}=\frac{q}{2\mu_{\chi 1}}$.

\begin{figure}[t!]
\centering
Comparison of Event Rates\\
\vspace{5pt}
\includegraphics[width=0.49\textwidth]{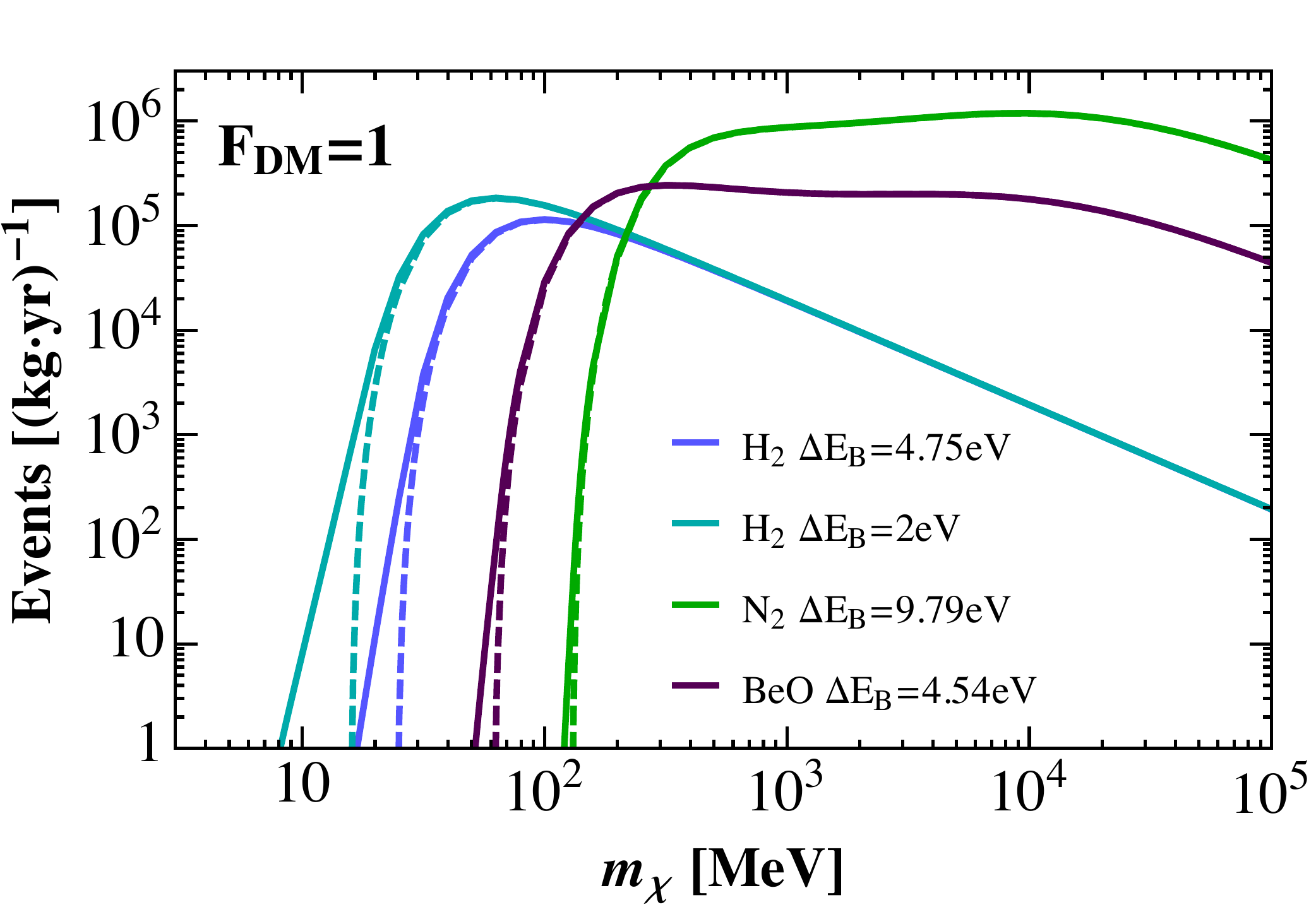} 
\hspace{0.01cm}
\includegraphics[width=0.49\textwidth]{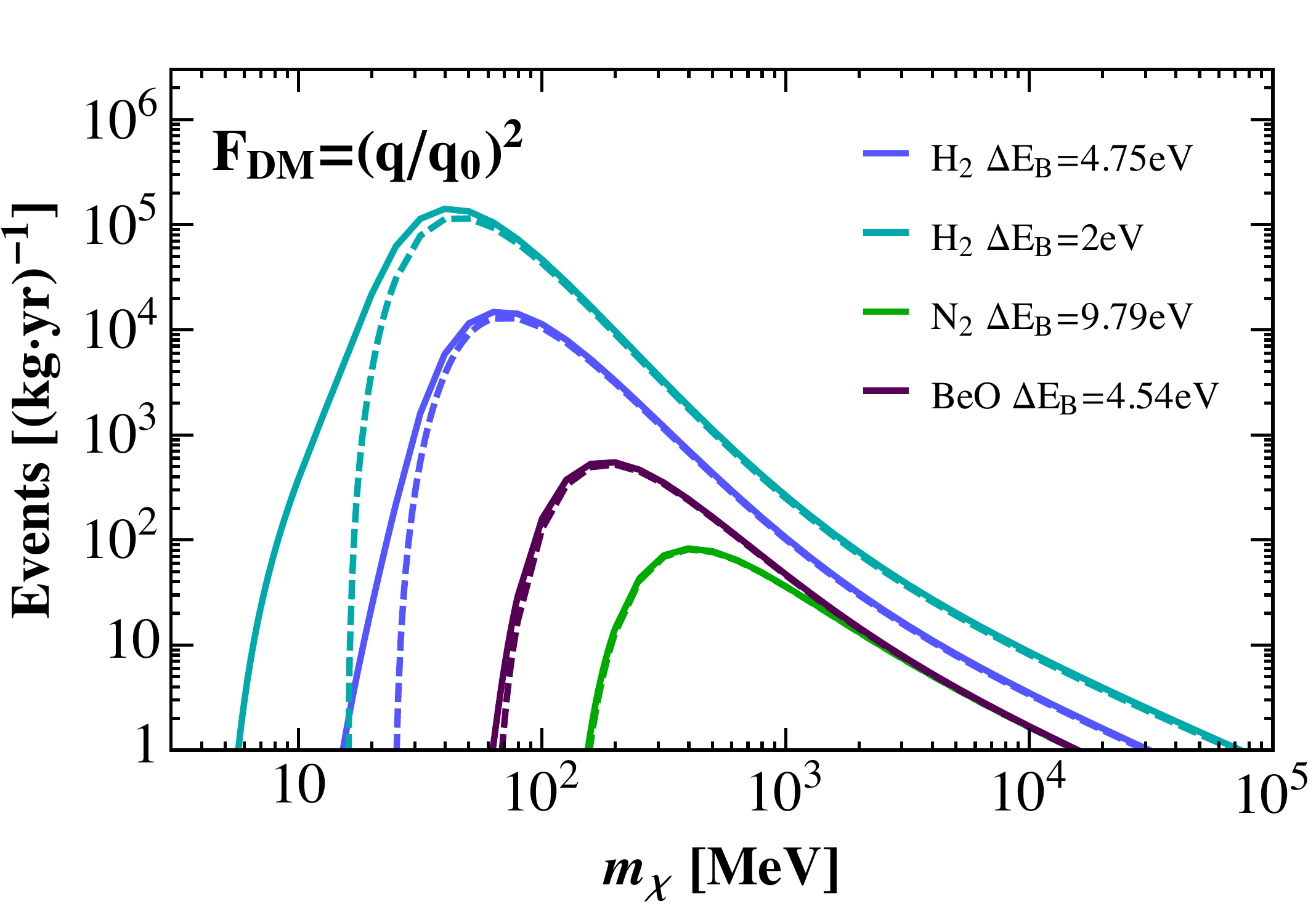} 
\caption{
Comparison of expected event rates for dissociation of  diatomic molecules:  H$_2$ with $\DE=4.75 \text{ eV}$ (\textit{blue}), an H$_2$-like molecule with $\DE=2 \text{ eV}$ (\textit{cyan}) ,  N$_2$ with $\DE=9.79 \text{ eV}$ (\textit{green}), and BeO with $\Delta E_B=4.54 \text{ eV}$ (\textit{purple}). For the case of BeO, the scattering is with the Be nucleus only. The rates are calculated both classically ({\it dashed lines}) and with quantum effects included ({\it solid lines}) for a DM-nucleon cross section $\sigmap = 10^{-37} \text{ cm}^2$, and for a DM form-factor of $F_{\rm DM} = 1$ ({\it left}) and $F_{\rm DM} \propto 1/q^2$ ({\it right}). The quantum rates are calculated using the Improved Born Approximation (see Sec.~\ref{sec:rates-sph-symm}).
\label{Fig:Rate_Compare}
}
\end{figure}

In the quantum regime the dissociation form factor becomes important. Combining these definitions with the equation for the scattering rate leads to the differential dissociation rate,
\beq
\frac{dR}{d \ln E_r} = \left[ f_PZ+f_N(A-Z) \right]^2 N_T \frac{\rho_{\chi}}{m_{\chi}}\frac{\sigmap}{8\mu_{\chi n}^{2}} \int d \ln E_{\rm int} \times q^2 \FDM \Fdis \eta(v_{\rm min})\,,
\label{eq:Rate_Quantum}
\eeq
with $v_{\rm min}$ taken from Eq.~(\ref{eq:vmin_General}).
The classical and quantum rates for various molecules, calculated in the classical approximation and in the Improved Born Approximation are shown in Fig.~\ref{Fig:Rate_Compare}. For BeO, we plot the interaction rate of DM with the Be nuclei. The left and right panels present the total rates for a range of DM masses with two DM form factors, $\FDM = 1$ and $\FDM = \left(\frac{q_0}{q}\right)^4$, respectively. 
For large DM masses, the approximations all coincide with the full result. However, at small DM masses the classical approximation breaks down, underestimating the true scattering rate. 

\begin{figure*}
\begin{center}
Quantum Corrections to Differential Event Rates\\
\vspace{5pt}
\includegraphics[width=0.49\textwidth]{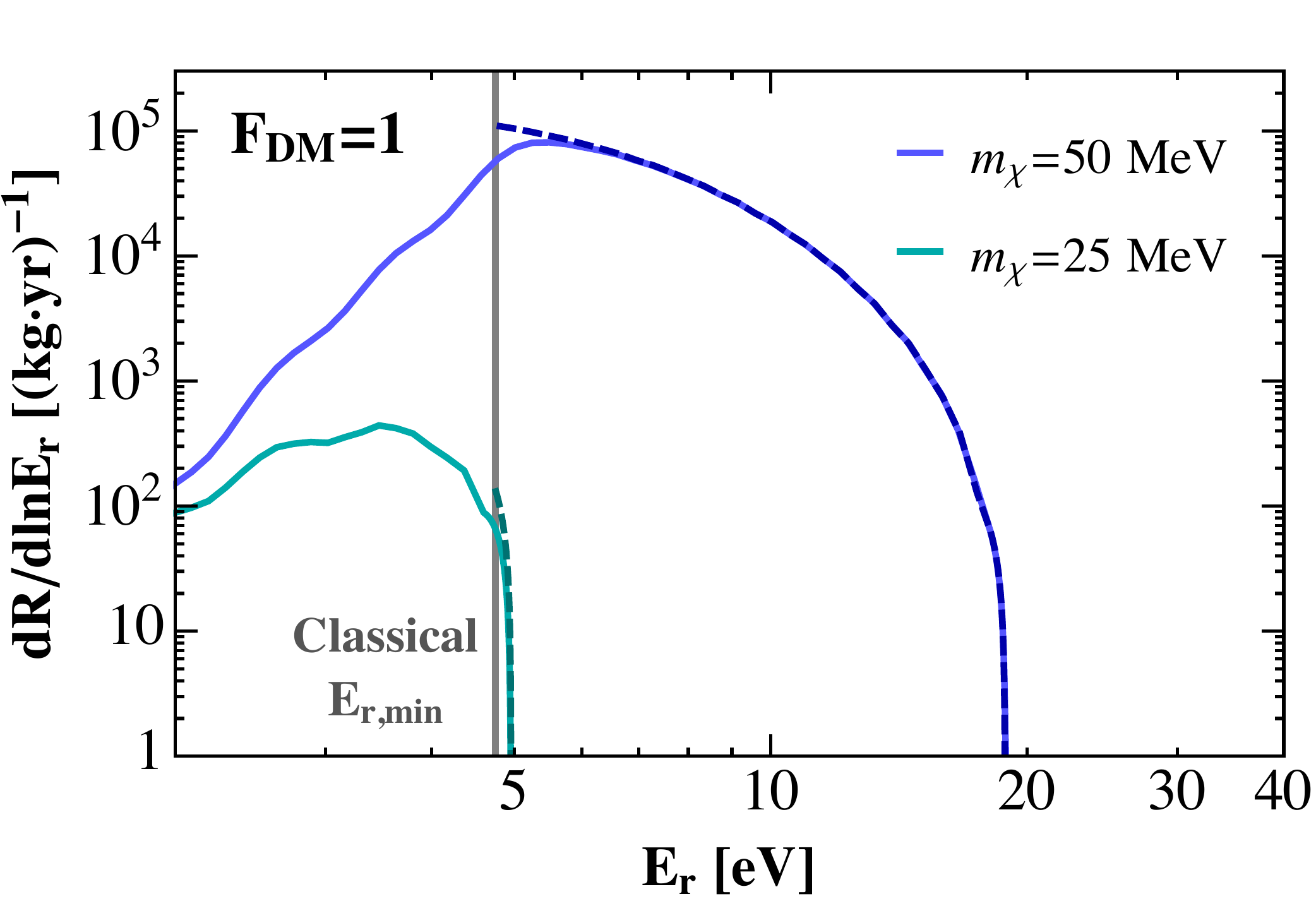}
\hspace{0.01cm}
\includegraphics[width=0.49\textwidth]{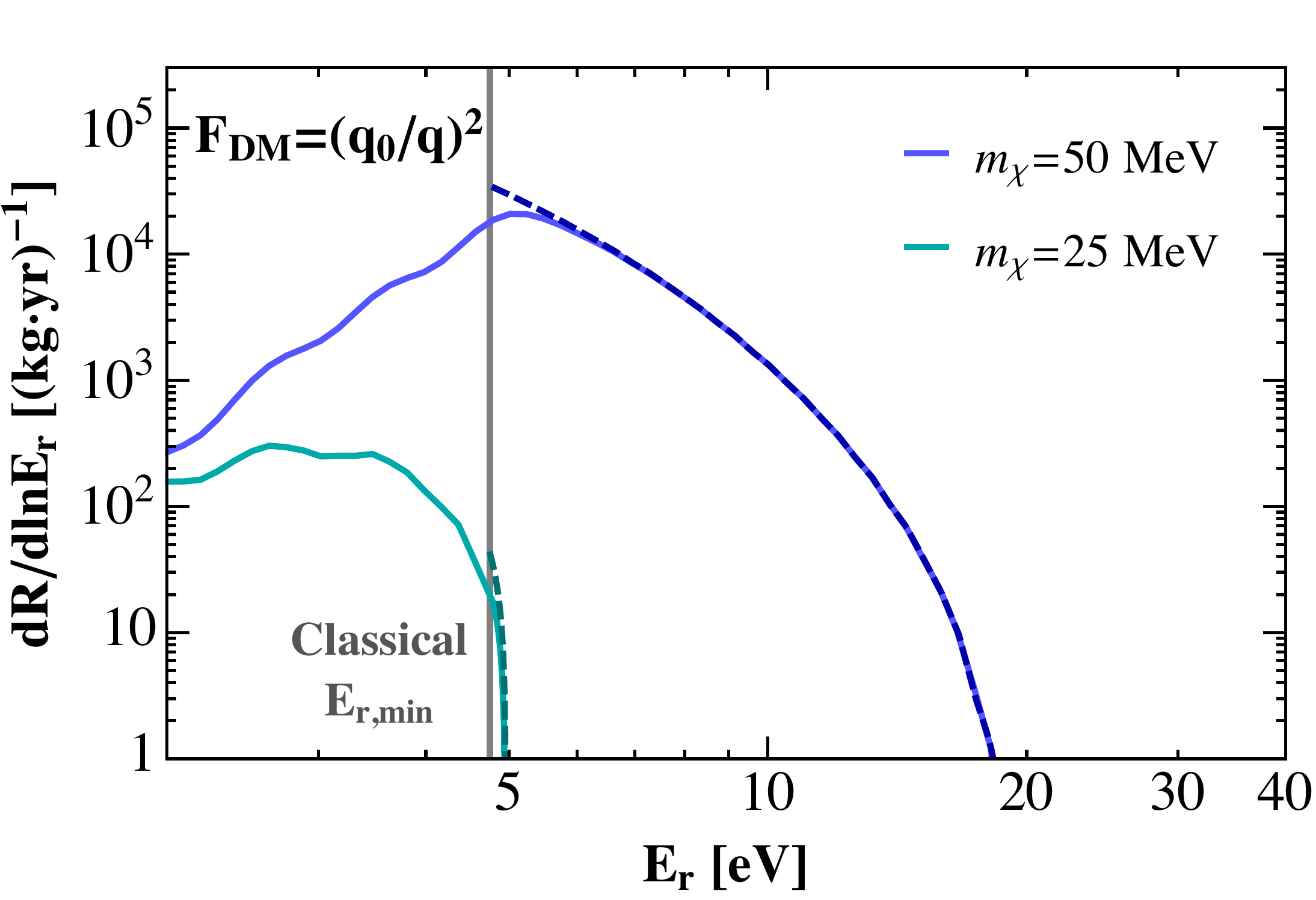} 
\caption{
Comparison of differential event rates in an H$_2$ molecule of the classical approximation (\textit{dashed lines}) with 
those including quantum effects (\textit{solid lines}). 
Gray vertical lines indicate the minimal classical recoil energy, $E_{r,\rm min} = \frac{m_1}{m_2}\DE$, which corresponds classically to $E_{\rm int}=0$. 
Quantum effects are important below the classical threshold, especially for lower DM masses, 
as illustrated by the {\it cyan} ($m_\chi = 25$~MeV) and {\it blue} ($m_\chi = 50$~MeV) curves.  
We set $F_{\rm DM} = 1$ ({\it left}) and  $F_{\rm DM} \sim 1/q^2$ ({\it right}). 
The quantum rates are calculated in the Improved Born Approximation (see Sec.~\ref{sec:rates-sph-symm}).
\label{Fig:dRdEr}
}
\end{center}
\end{figure*}

The reason for the difference between the Improved Born Approximation and classical calculations can be seen in Fig.~\ref{Fig:dRdEr}.
Classically, there is a minimal value for the recoil energy, $E_{r,min} = \frac{m_1}{m_2} \Delta E_B$, whereas for the quantum calculation there is no such rigid minimal value (see Sec.~\ref{sec:quantum-effects}). 
Thus, the true quantum process receives contributions from lower values of recoil energy than in the classical approximation. 
This correction is important for low DM masses. 
Below a certain DM mass, there is no classical contribution, and the entire measurable rate is due to quantum effects. This is evident in the figure, where the classical contribution is almost negligible compared to the quantum rate for the lower DM mass. 
These effects depend largely on the tail of the velocity profile and are therefore extremely sensitive to the exact shape of this tail.

We have checked that the Improved Born Approximation agrees very well with the full calculation (see also Fig.~\ref{Fig:FF_Comparison} and discussion in Sec.~\ref{sec:rates-sph-symm}), even for DM masses close to the dissociation threshold. We stress that the classical rates, which are straightforward to compute, can be safely used for masses slightly above threshold.  In the left panel of Fig.~\ref{Fig:Rate_Compare}, the event rate for molecular nitrogen and for beryllium oxide exhibit non-trivial features in the mass range $m_\chi \approx 0.1-10 \text{ GeV}$. These effects can be understood when considering the three mass scales involved in the process: (i) the minimal DM mass for which dissociation occurs, (ii) the mass of the nucleon ($\sim1 \text{ GeV}$), and (iii) the mass of the nucleus. In particular, the integral over $\eta(v_{min})$ depends on all three of these scales, while the reduced mass, $\mu_{\chi n}$, turns over at approximately the nucleon mass. For molecular nitrogen, these masses are separated by about one decade each and for beryllium slightly more, accounting for the features visible in the event rate. The result depends on $\mu_{\chi n}$, since we have chosen a fixed nucleon cross section. For the non-trivial DM form factor, $\FDM \propto \frac{1}{q^4}$, this effect is washed out.

\begin{figure}[!htb]
\centering
\vspace{0.1cm}
{\small{H$_2$-like Molecule}}\\
\includegraphics[width=0.47\textwidth]{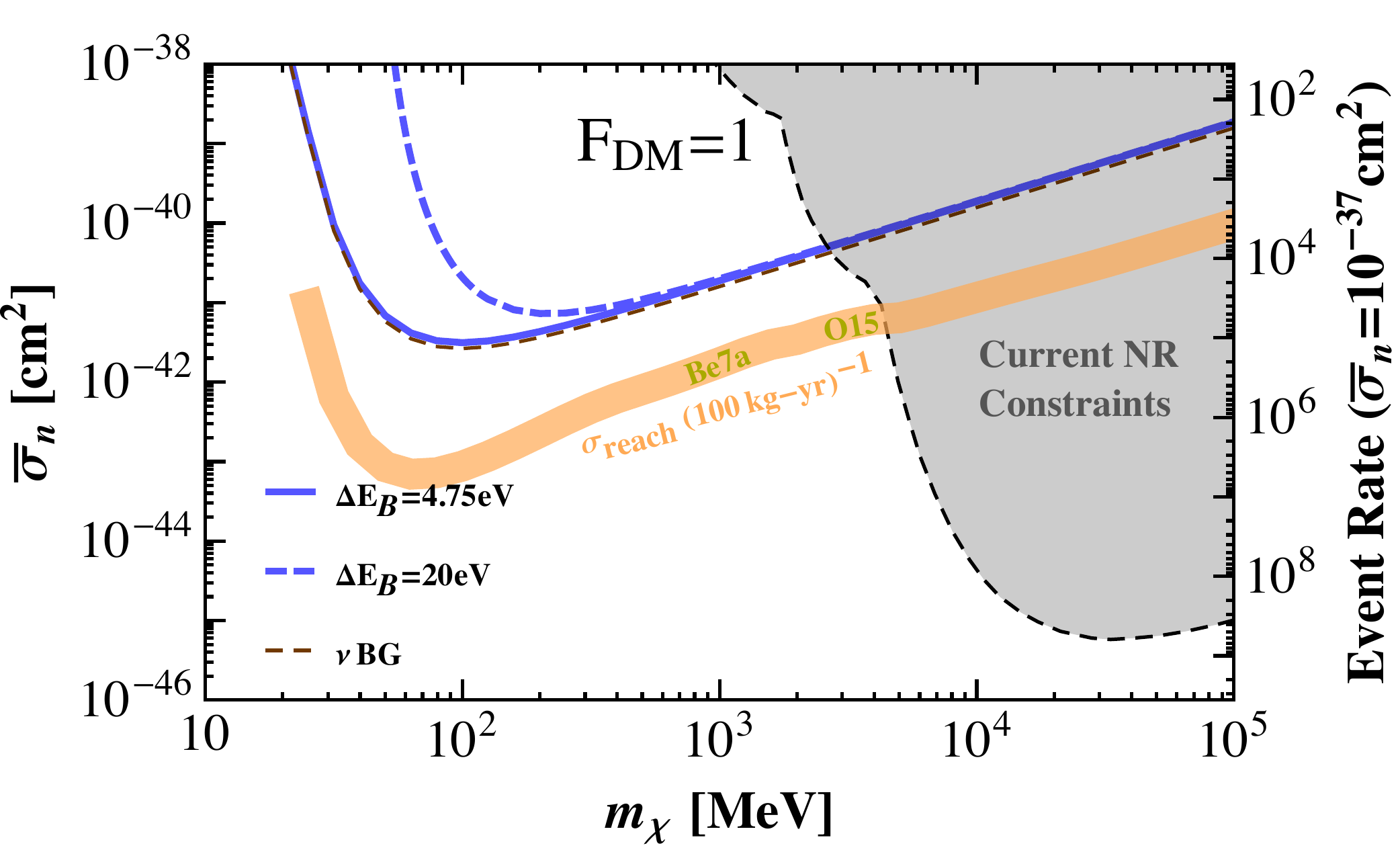}
~~
\includegraphics[width=0.47\textwidth]{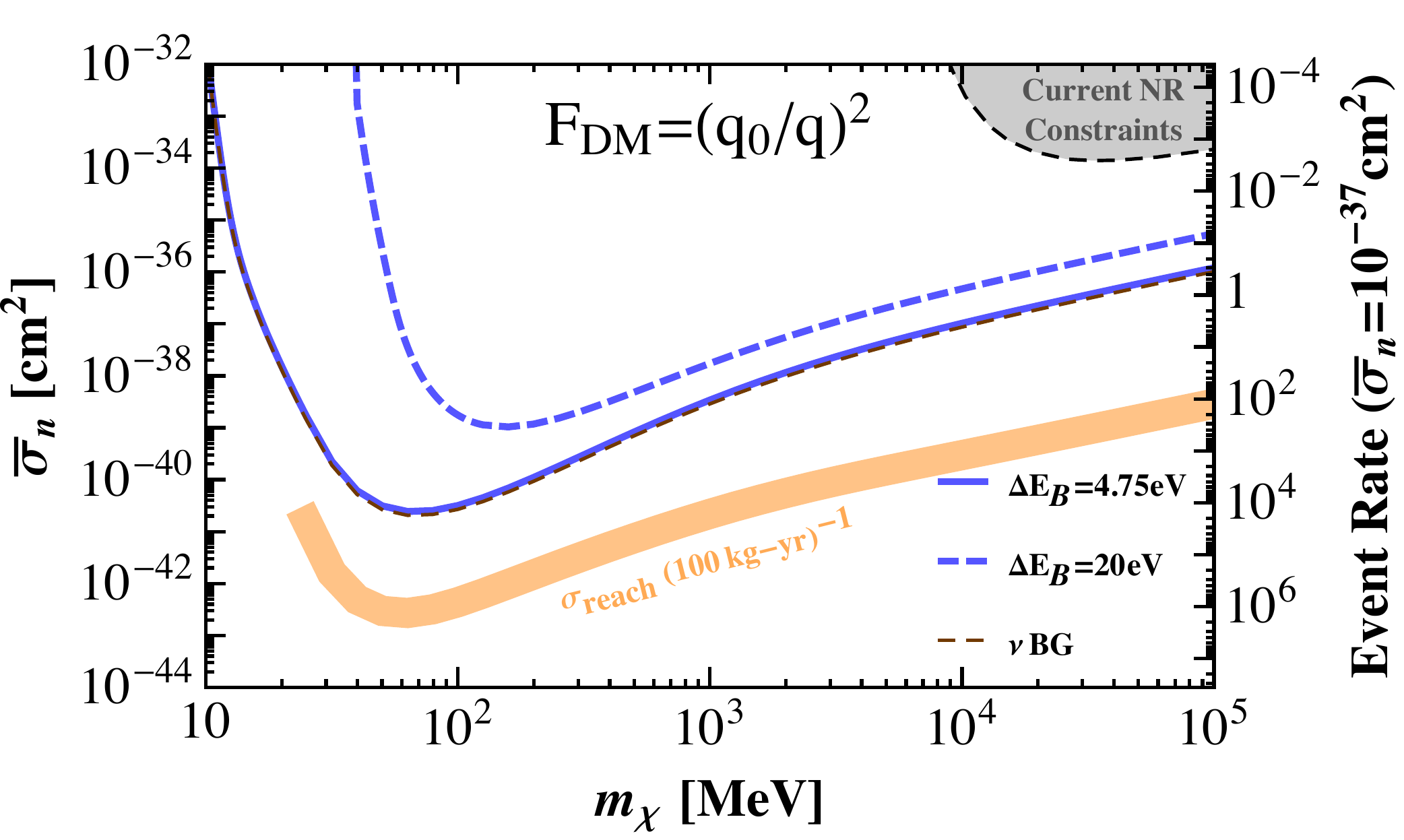}
\vspace{0.1cm}
{\small{N$_2$-like Molecule}}\\
\includegraphics[width=0.47\textwidth]{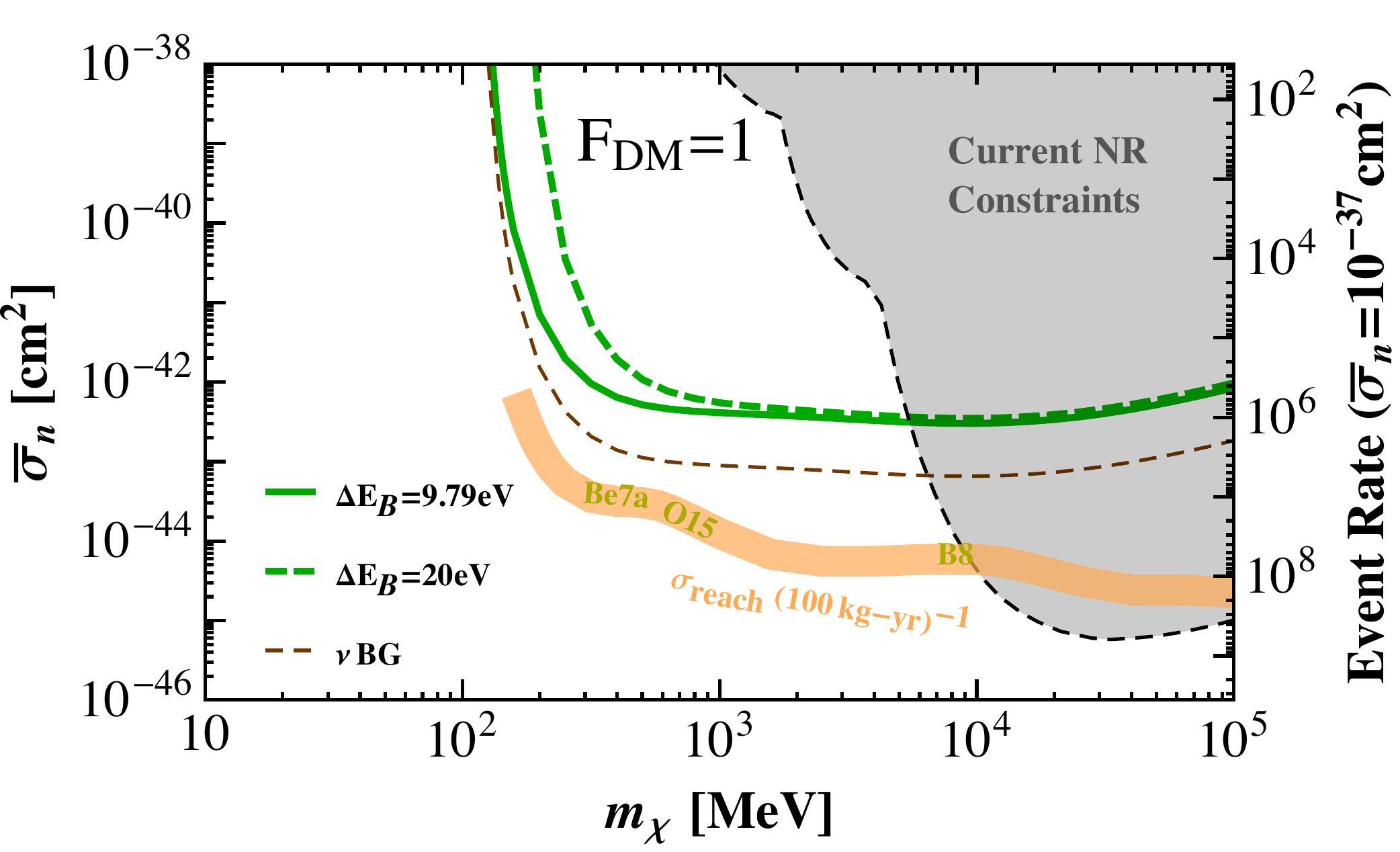}
~~
\includegraphics[width=0.47\textwidth]{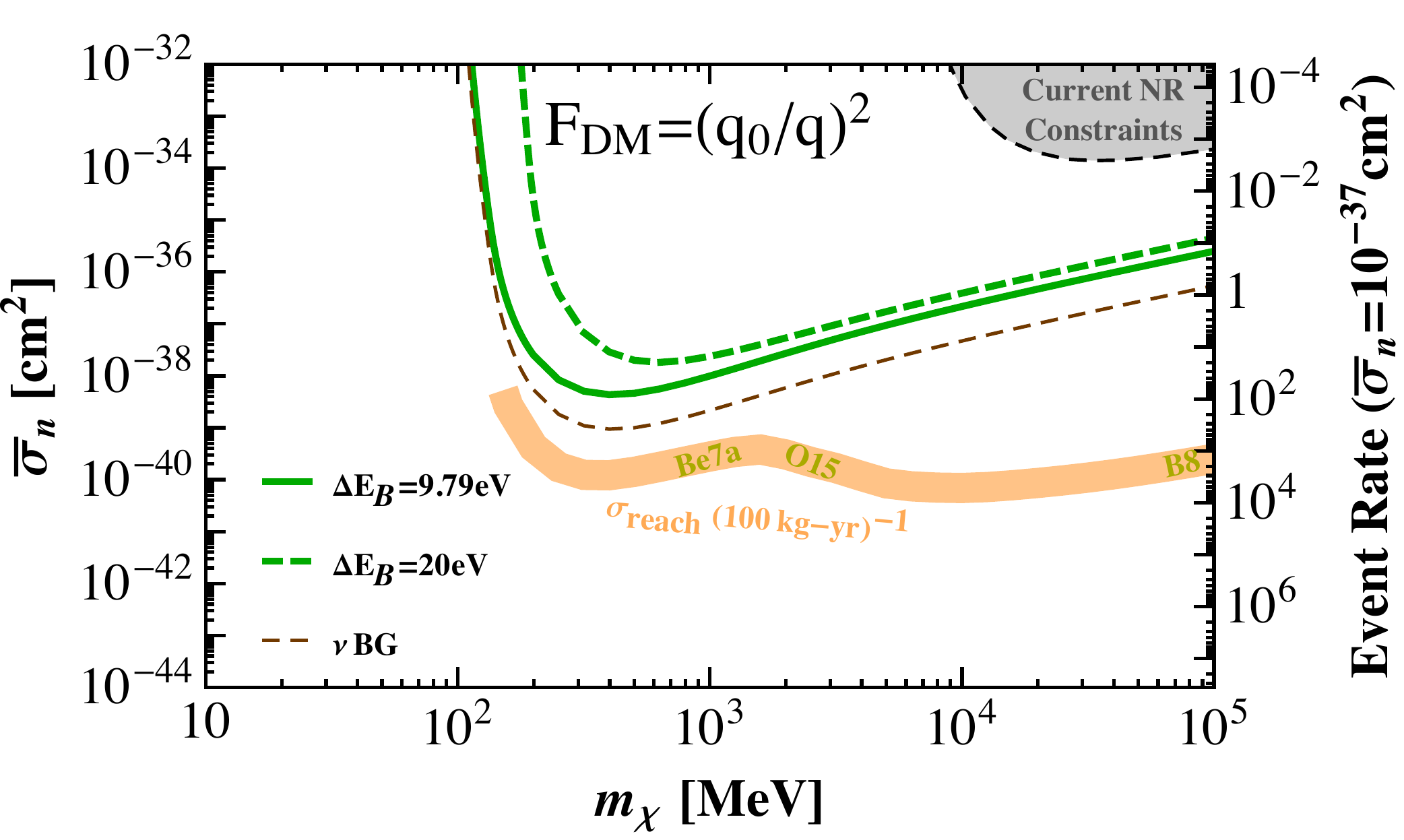}
\vspace{0.1cm} 
{\small{BeO-like Molecule}}\\
\includegraphics[width=0.47\textwidth]{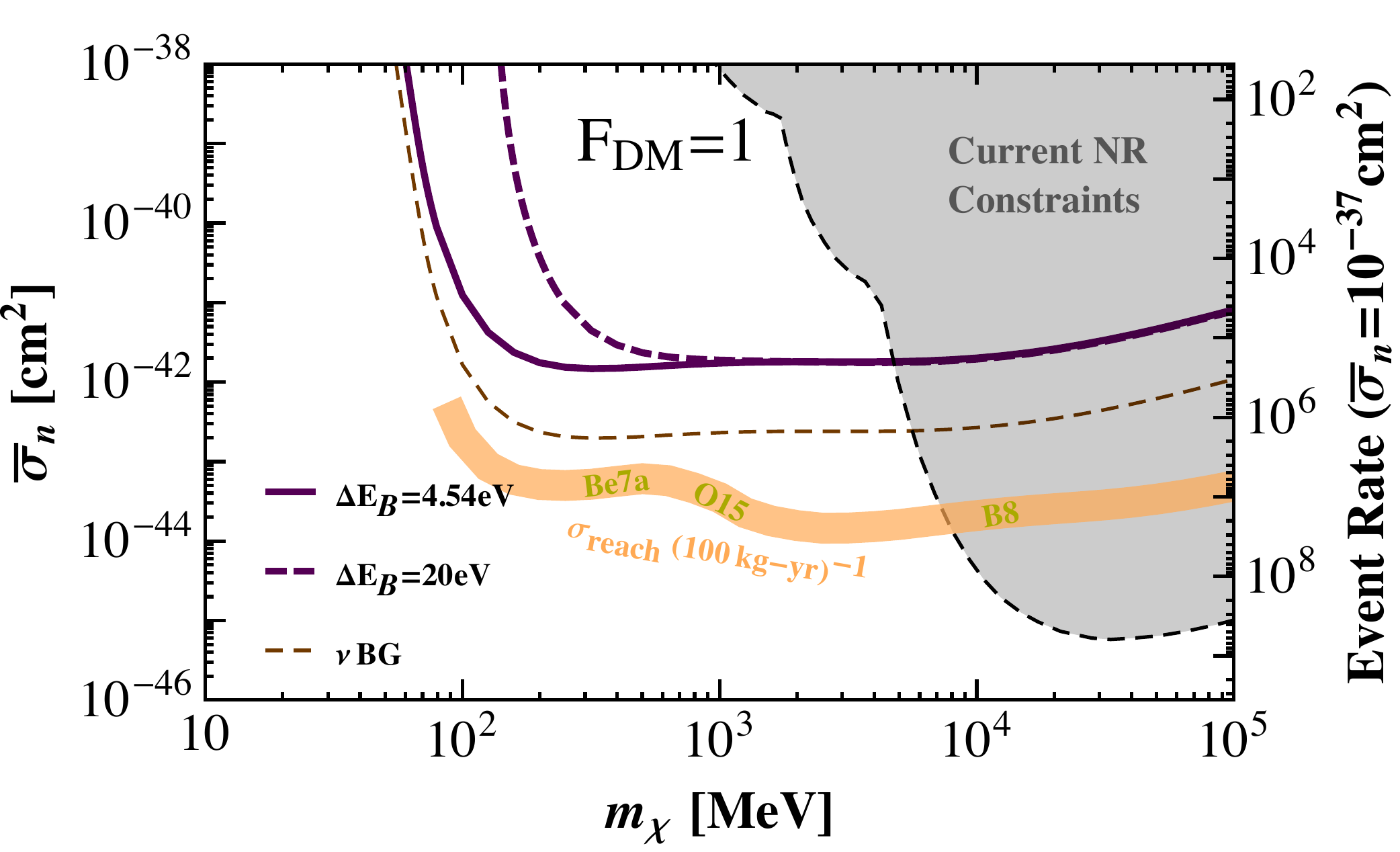}
~~
\includegraphics[width=0.47\textwidth]{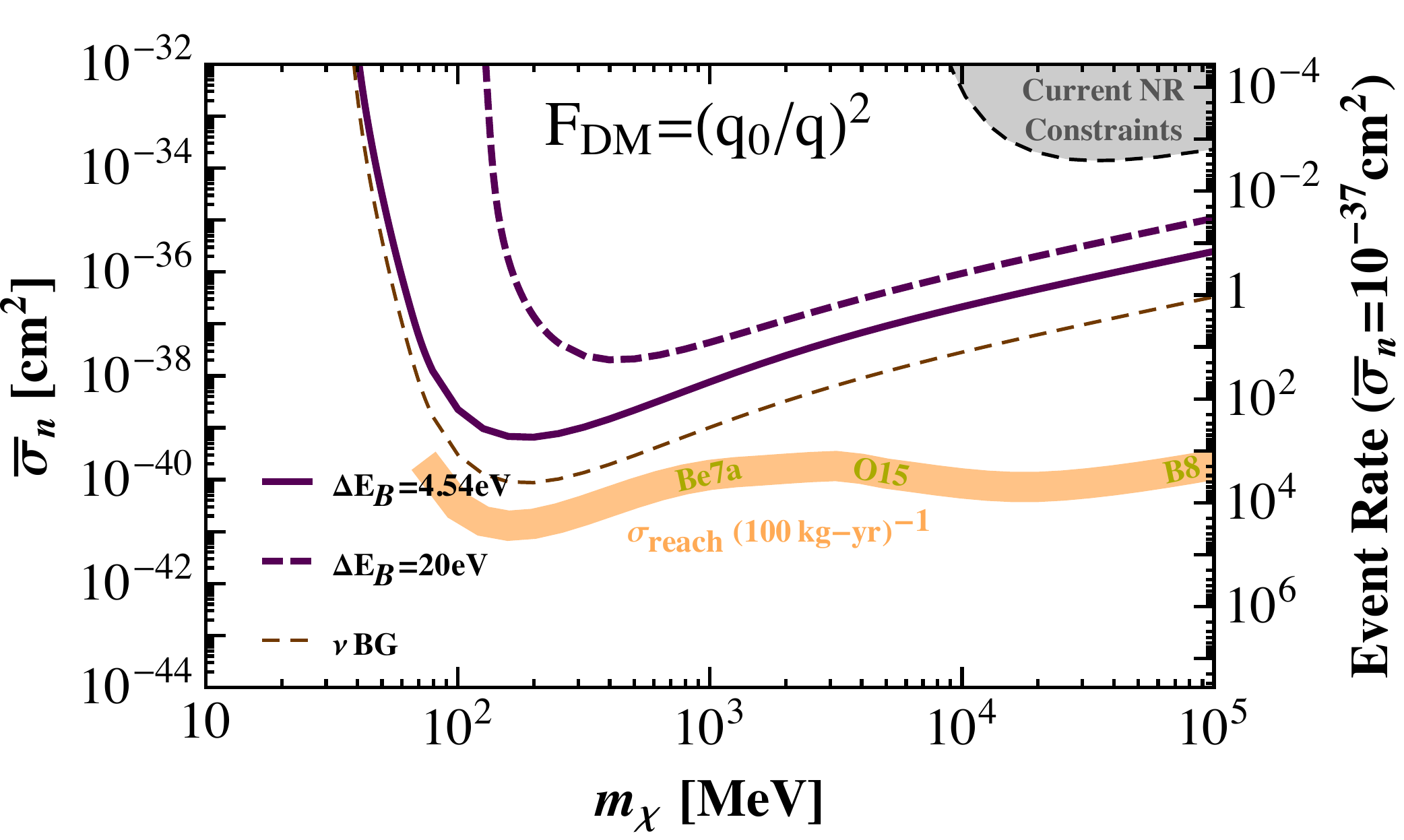}
\caption{
Potential cross-section sensitivities of a background-free search for chemical bond breaking with 1\,kg$\cdot$year exposures (left axes). 
Equivalently, the right axes show the event rate assuming $\sigmap = 10^{-37}\,\text{cm}^2$. 
{\it Top}, {\it middle}, and {\it bottom} plots correspond to H$_2$, N$_2$, and BeO-like targets, respectively, with solid curves calculated using the true binding energy for each target, and dashed curves with the binding energy set to 20\,eV, which is closer to the energy barrier for many crystal transitions (these are illustrative and are not intended to represent realistic experiments). For the case of BeO the result is shown for DM scattering with Be only.
Quantum effects are included for all curves using the Improved Born Approximation. 
{\it Left} and {\it right} plots correspond to $F_{\rm DM} = 1$ and $F_{\rm DM} \sim 1/q^2$, respectively. 
Black dashed lines/gray shaded regions show current nuclear recoil bounds (see text). 
%from CRESST II, CDMSLite, SuperCDMS, and LUX.
Dark orange dashed lines show where the solar neutrino event rate equals the DM rate: reaching below these would require background subtraction. 
Orange bands indicate the ultimate reach using background subtraction, limited only by statistics and theoretical uncertainties in solar neutrino modeling, assuming a 100\,kg$\cdot$year exposure.
}
\label{Fig:Sens_Plots}
\end{figure}

The potential experimental sensitivities, assuming zero background, are shown in Fig.~\ref{Fig:Sens_Plots} for two DM form factors, $\FDM = 1$ and $\FDM = \left( \frac{q_0}{q} \right)^4$. In each panel, the left axis shows the cross-section sensitivity at $95\%$ confidence level (corresponding to 3.6 observed signal events) for an experiment with 1\,kg$\cdot$year exposure and sensitivity to a single bond-breaking event. Equivalently, the right axis shows the event rate assuming a cross section of $\sigmap = 10^{-37} \text{ cm}^2$. Solid lines correspond to the target material being molecular hydrogen, molecular nitrogen, and beryllium oxide, in the top, middle, and bottom plots, respectively. We also show the results for molecular hydrogen, nitrogen, and beryllium oxide-like molecules bound by a potential with binding energy $\DE = 20 \text{ eV}$, which is in the range of typical binding energies of atoms within crystals. 
Since the other parameters of the Morse potential do not have a significant effect on the expected rate, for the larger binding energies the parameters $r_{0}$ and $\alpha$ have been chosen to be equal to those of the real molecules. 
The \textit{dashed dark orange} lines in all panels of Fig.~\ref{Fig:Sens_Plots} correspond to the cross section for which the expected number of DM events is equal to the expected number of neutrino events (see a detailed discussion regarding solar neutrino detection and background in Sec.~\ref{sec:Neutrinos}). This limit is exposure independent (since both the DM rate and the neutrino rate scale linearly with exposure) and represents the approximate cross section for which neutrino background reduction becomes important.

For cross sections on and below the dashed brown lines in Fig.~\ref{Fig:Sens_Plots}, a likelihood analysis based on modeling of the solar neutrino event rate and spectrum may be able differentiate between a neutrino-only hypothesis and a neutrino + DM hypothesis. Details regarding such an analysis are presented in Sec.~\ref{sec:Neutrinos}. The thick orange bands in Fig.~\ref{Fig:Sens_Plots} correspond to the maximal cross section that can be probed after such a neutrino background subtraction for a 100 kg$\cdot$year exposure. Features present in these curves for certain DM masses arise if the DM recoil spectrum for that mass mimics one or more families of the solar-neutrino spectra. In such a case, the uncertainty in the neutrino flux dominates and neutrino subtraction becomes less efficient. The neutrino family that dominates the uncertainty is shown in Fig.~\ref{Fig:Sens_Plots} for various DM masses. Also shown in the figure are the bounds derived from current nuclear recoil experiments, namely CRESST II~\cite{Angloher:2014myn}, CDMSLite~\cite{Agnese:2013jaa}, SuperCDMS~\cite{Agnese:2014aze}, and LUX~\cite{Akerib:2015rjg}.

\subsection{Annual Modulation}
To illustrate the prospects of detection for various setups, we present a plot of the discovery potential as a function of the number of background events. We define the modulation signal as,
\beq
\Delta S_{\rm mod} \equiv \frac{1}{2}\left[ R_{\rm max} - R_{\rm min} \right]\,,
\eeq
where $R_{\rm max}$, $R_{\rm min}$ are the maximal and minimal rates expected from modulation of the Earth's velocity with respect to the DM halo~\cite{Savage:2006qr}. 
For a given DM mass and target molecule, we calculate the cross section for which
\beq
\frac{\Delta S_{\rm mod}}{\sqrt{S + B}} = 5,
\eeq
where $S$ is the total number of events expected from DM interactions and $B$ is the number of background events.

\begin{figure*}
\centering
\includegraphics[width=0.5\textwidth]{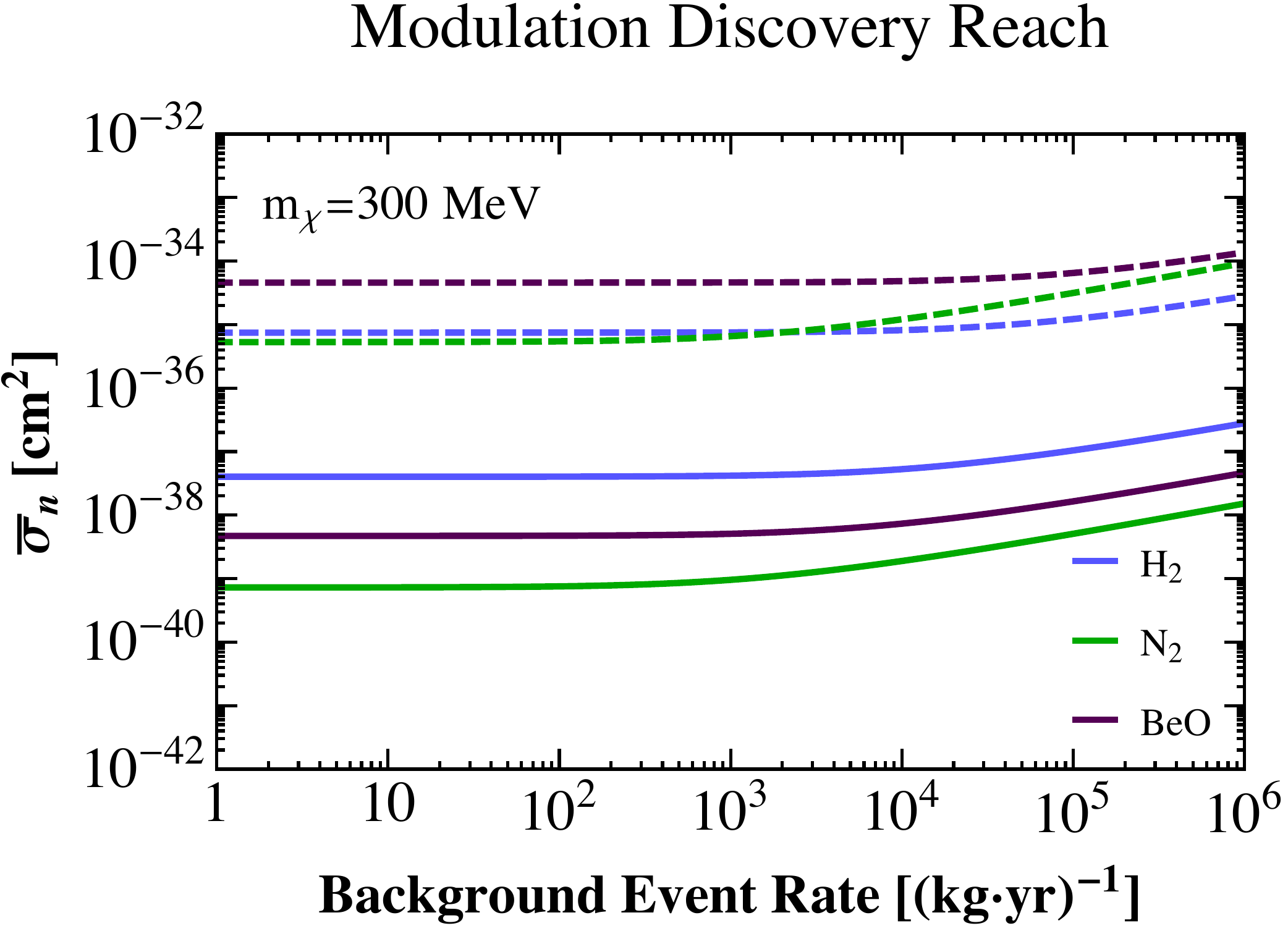}
\hfill
\includegraphics[width=0.49\textwidth]{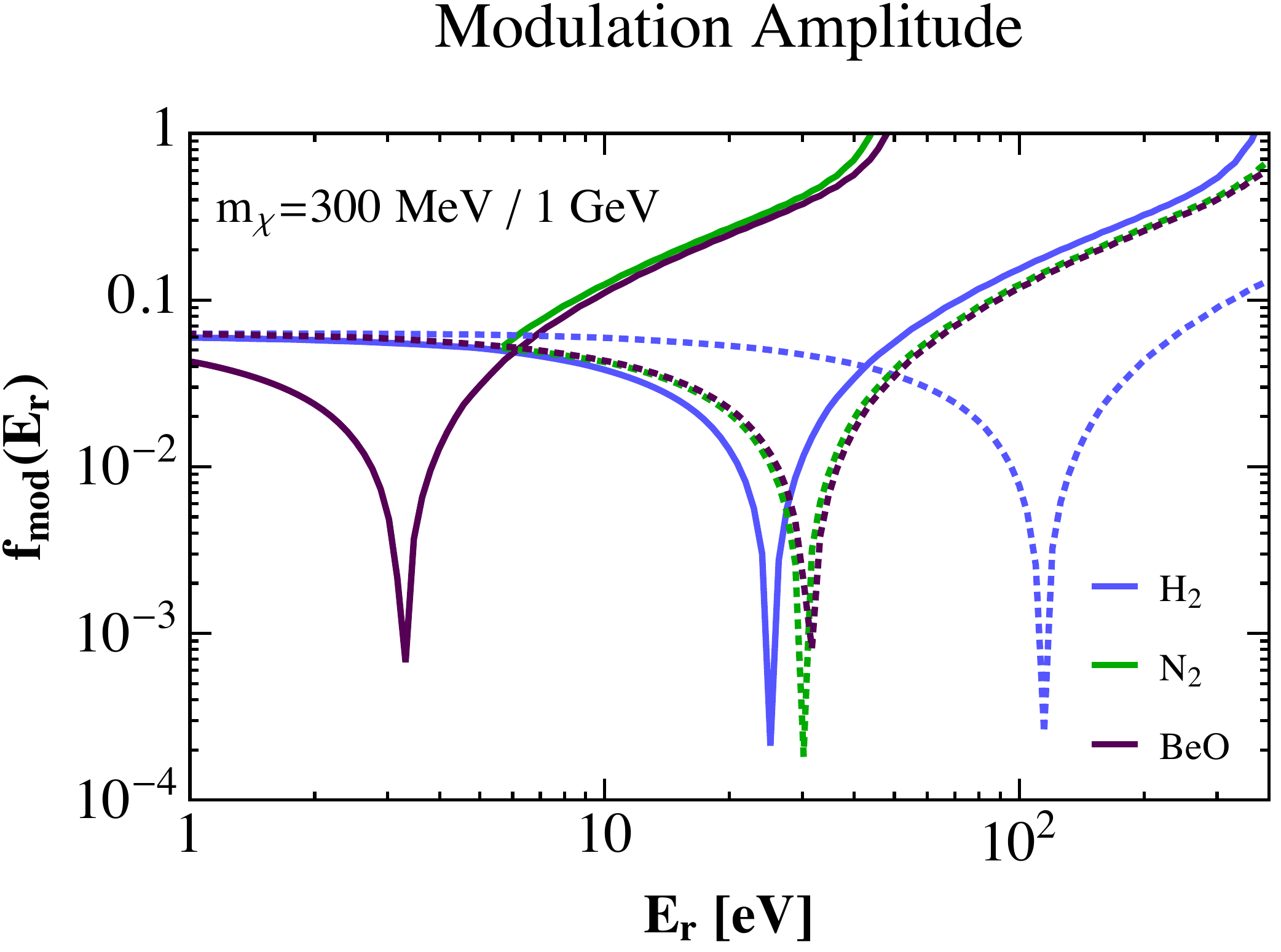}
\caption{{\it Left:}  The discovery reach as a function of the background event rate using annual modulation, for three illustrative diatomic molecules H$_2$ (\textit{blue}), N$_2$ (\textit{green}), and BeO (\textit{purple}), for a DM mass $m_\chi = 300 \text{ MeV}$. Solid (dashed) lines correspond to a DM form-factor of $F_{\rm DM} = 1$ ($F_{\rm DM} \sim 1/q^2$). 
{\it Right:} Modulation amplitude, $f_{\text{mod}}(E_r)$ as defined in Eq.~(\ref{eq:Mod_Amp}), as a function of recoil energy, for 
$F_{\rm DM} = 1$, and for $m_\chi = 300 \text{ MeV}$ ({\it solid lines}) and $m_\chi = 1 \text{ GeV}$ ({\it dashed lines}). 
The minimum of each curve occurs at the value of $E_r$ for which the phase of modulation reverses. Quantum effects are included for all curves using the Improved Born Approximation.   (Sub-)daily modulation for a non-spherically symmetric target, which may significantly improve the discovery reach, is not shown. 
\label{fig:Mod_Reach}
}
\end{figure*}
In the left panel of Fig.~\ref{fig:Mod_Reach}, we show the discovery reach for H$_2$, N$_2$, and BeO, for a 300 MeV DM mass and for 
two DM form factors. In the left panel, we show the modulation amplitude, $f_{\rm mod}$, as a function of the recoil energy, 
which is defined by,
\beq
f_{\rm mod}(E_r) \equiv \frac{\partial \Delta S_{\rm mod}(E_r)/\partial E_r}{\partial S(E_r)/\partial E_r}.
\label{eq:Mod_Amp}
\eeq
The features in these curves corresponds to the values of $E_r$ for which the phase of modulation reverses (see \cite{Savage:2006qr} for details). A more complex analysis is required for non-spherically-symmetric targets, which can also produce a (sub-)daily event-rate modulation that depends on the orientation of the target with respect to the DM ``wind''. 

\subsection{Example: Light Dark Matter coupled to a Dark Photon}

Until now we have not specified any model for the DM.  For concreteness, we briefly discuss a model for which the dark sector includes a 
fermionic DM particle charged under a local $U(1)$ symmetry with gauge coupling $g_D$ (we denote $\alpha_D\equiv g_D^2/4\pi$)~\cite{Essig:2011nj,Essig:2015cda}. The dark photon couples to the SM photon via kinetic mixing (with the parameter denoted as $\epsilon$) 
and consequently, the DM form factor and $\sigmap$ are given by
\begin{eqnarray}
\FDM & = & \frac{(M_{A_D}^2 + q_0^2)^2}{(M_{A_D}^2 + q^2)^2} =
\begin{cases}
1 ,& M_{A_D} \gg q_0 \\
\left(\frac{q_0}{q}\right)^4 ,& M_{A_D} \ll q_0 \\
\end{cases}\,,
\label{eq:DD_FF_2} \\
\sigmap & = & \frac{16 \pi  \alpha \alpha_D \epsilon^2 \mu_{\chi n}^2}{(M_{A_D}^2 + q_0^2)^2} =
\begin{cases}
\frac{16 \pi \alpha \alpha_D \epsilon^2 \mu_{\chi n}^2}{M_{A_D}^4} ,& M_{A_D} \gg q_0 \\
\frac{16 \pi \alpha \alpha_D \epsilon^2 \mu_{\chi n}^2}{q_0^4} ,& M_{A_D} \ll q_0 \\
\end{cases} \,.
\label{eq:sigma_p_2}
\end{eqnarray}

\begin{figure*}
\centering
\includegraphics[width=0.49\textwidth]{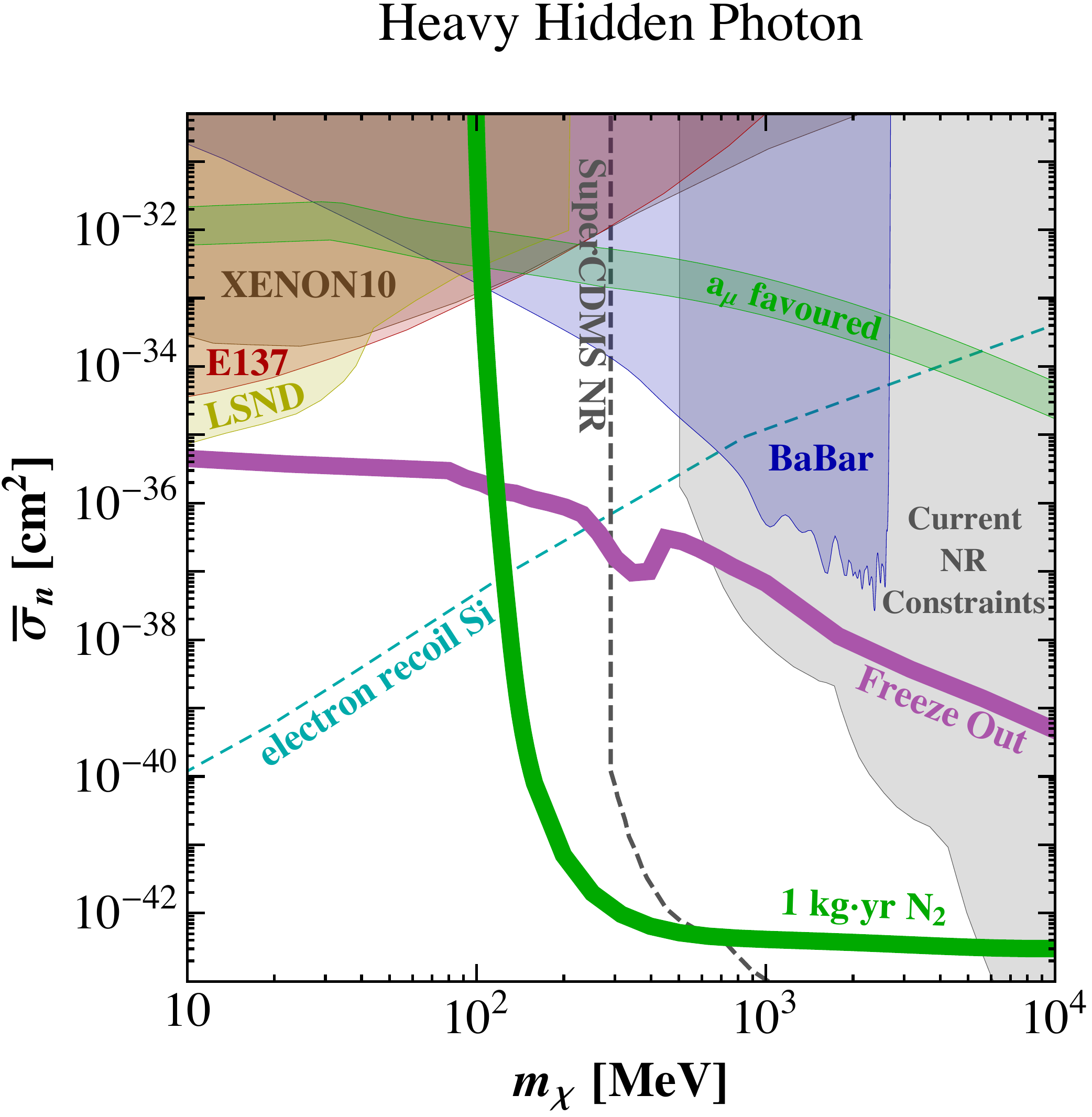}
\hspace{0.0cm}
\includegraphics[width=0.49\textwidth]{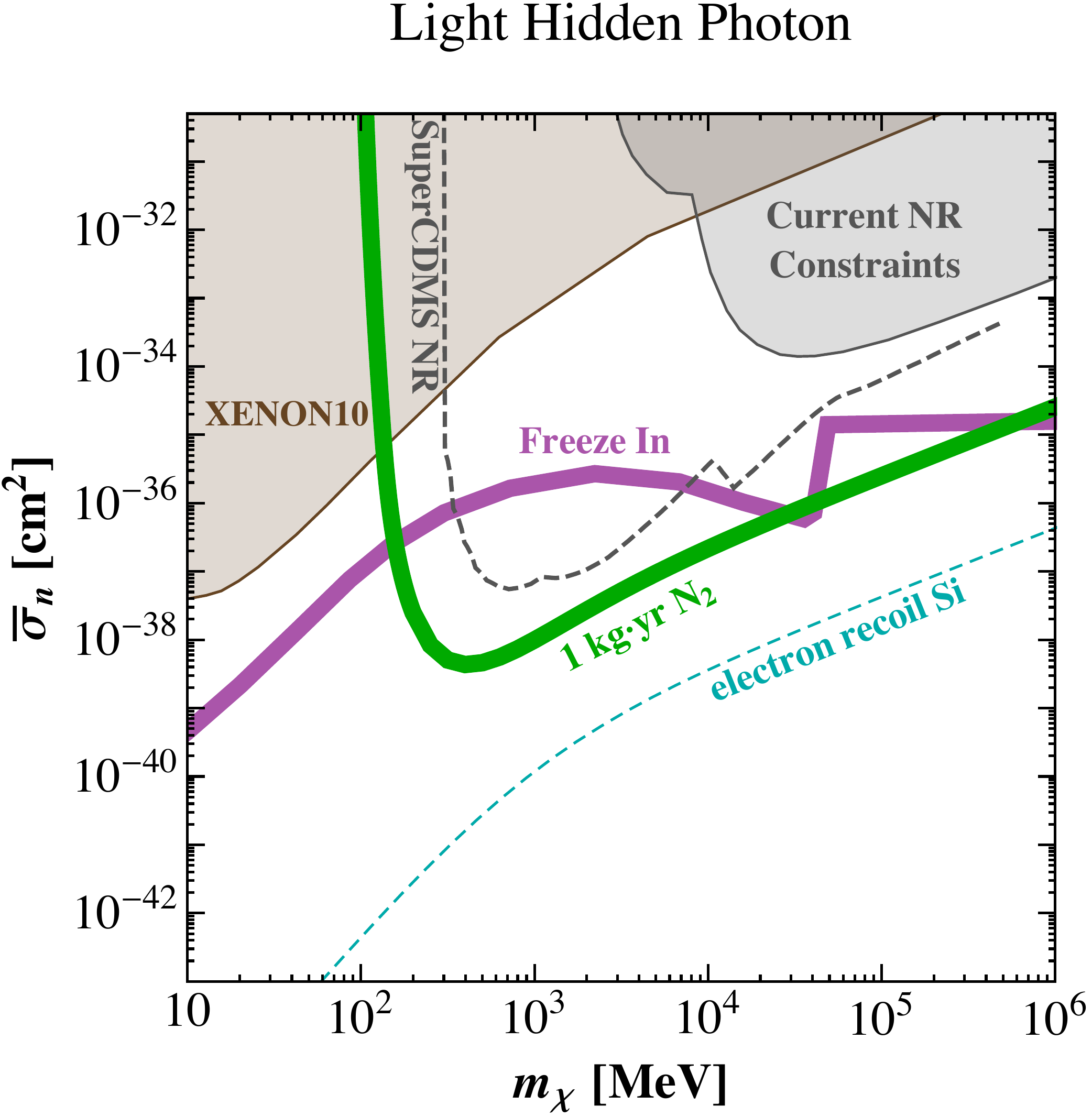}
\caption{
Potential reach, and existing experimental bounds, for a dark-photon model in two regimes: a dark photon heavier than the DM ({\bf left})  and an ultra-light dark photon ({\bf right}). For the former we take  $M_{A_D} = 3 m_\chi$, while for the latter we take $M_{A_D} \ll q_0 = 100$ keV. 
Shaded regions show existing constraints from electron-recoil analysis of XENON10 data (\textit{brown}), electron beam-dump E137 (\textit{red}), proton beam-dump LSND (\textit{yellow}), BaBar search for $e^+e^-\rightarrow \gamma+\textit{invisible}$ (\textit{blue}), conventional nuclear recoil searches CRESST II, CDMSLite, SuperCDMS, and LUX (labeled Current NR Constraints) (\textit{black}) and the $2\sigma$ region that can explain the muon anomalous magnetic moment, $a_\mu$ (\textit{light green}). Also shown are cross sections for which one gets the correct relic abundance (\textit{thick purple}) via freeze-out (heavy hidden photon) or freeze-in (light hidden photon). The maximal future sensitivity is shown for DM scattering off electrons in a 
silicon target at SuperCDMS with a 10~kg$\cdot$years exposure on silicon (\textit{dashed cyan}).  Also shown is the SuperCDMS SNOLAB nuclear recoil projection (\textit{dashed black}). 
The sensitivity of a hypothetical molecular N$_2$-based detector with exposures of 1~kg-year are shown in thick green. 
}
\label{Fig:Exp_Bounds}
\end{figure*}

Current bounds from a range of experimental data for the two limiting cases $M_{A_D} \gg q_0$ and $M_{A_D} \ll q_0$ are presented in Fig.~\ref{Fig:Exp_Bounds} where for concreteness we take $M_{A_D} = 3 m_\chi$ for the heavy dark photon regime and $M_{A_D} \ll q_0$ for the light dark photon regime. Presented in both panels are nuclear-recoil cross-section constraints from a number of experiments. For cases in which electron recoil measurements constrain the dark photon parameter space, the results have been recast to constrain the DM-nucleon cross section expected from the same set of parameters. The constraints presented in the figure include self-interaction~\cite{Tulin:2013teo,Randall:2007ph} and unitarity bounds~\cite{Davoudiasl:2015hxa} on $\alpha_D$, as well as bounds from  
electron recoil analysis of XENON10 data (\textit{brown})~\cite{Essig:2012yx},  electron beam-dump E137
(\textit{red})~\cite{Batell:2014mga,Bjorken:1988as}, proton beam-dump LSND (\textit{yellow})~\cite{deNiverville:2011it, Batell:2009di} and BaBar search for $e^+e^-\rightarrow \gamma+\textit{invisible}$ (\textit{blue})~\cite{Batell:2009di}. 
We further show the  current nuclear recoil constraints from the conventional nuclear recoil searches  CRESST II~\cite{Angloher:2014myn}, CDMSLite~\cite{Agnese:2013jaa}, SuperCDMS~\cite{Agnese:2014aze}, and LUX~\cite{Akerib:2015rjg} (\textit{black}) 
and the $2\sigma$ region in parameter space for which a dark photon can explain the discrepancy between the measurement and the SM prediction for the muon anomalous magnetic moment, $a_\mu$ (\textit{light green})~\cite{Pospelov:2008zw}. In the left panel, the \textit{thick purple}   line shows the cross section for which the correct relic abundance is obtained via freeze-out of DM to SM particles through an off-shell dark photon (e.g.~\cite{Boehm:2003hm,Essig:2015cda,Izaguirre:2015yja}). Above or below this line, the model must be slightly revised in order to account for the observed relic abundance. In the right panel, the dark photon is effectively massless and the coupling of DM to the SM is so small that thermal equilibrium is never reached. In this scenario, the relic abundance can be obtained via freeze-in~\cite{Hall:2009bx} from $2\rightarrow2$ annihilation of SM particles to DM as well as Z boson decays to DM~\cite{Essig:2011nj,Chu:2011be}. The same colored line in this panel shows the parameter space for which the correct relic abundance is achieved. Dashed lines in both panels show 
the SuperCDMS SNOLAB nuclear recoil projection~\cite{Agnese:2015ywx}  (\textit{dashed black}) as well as the maximum future sensitivity from a SuperCDMS-like experiment searching for DM-electron scattering with a threshold of one electron and an exposure of 10~kg$\cdot$years~\cite{Essig:2015cda}   (\textit{dashed cyan}). 
The SuperCDMS NR projection for the light dark photon case has been simply rescaled from the $\FDM=1$ case, and neglects possible differences in the signal-versus-background shape. Finally, presented in the figure are potential cross-section sensitivities for a molecular $N_2$ based experiment with 1~kg$\cdot$year exposure. Such an experiment could be competitive to, and in some cases more sensitive than, the bounds and projections discussed above. Importantly, if the DM coupling with SM particles is leptophobic, then electron-recoil experimental constraints have no sensitivity, 
whereas a molecular based experiment would be sensitive to such a scenario. 

%%%%%%%%%%%%%%%%%%%%%%%%%%%
\section{Dissociation Rates for Solar Neutrinos} 
\label{sec:Neutrinos}
%%%%%%%%%%%%%%%%%%%%%%%%%%%

An experiment sensitive to nuclear interactions with weakly interacting particles and low threshold can be sensitive to solar neutrinos.  Solar neutrinos will be an irreducible background for DM searches with large exposures, but since much of the low-energy solar neutrino spectrum is yet to be fully measured, their detection is itself of great interest. In what follows, we discuss the bond-breaking event rate due to neutrino interactions and the corresponding irreducible background relevant for DM direct detection.

\subsection{Neutrino Scattering}
\label{sec:Neutrino_Scattering}

\begin{figure}[t]
\centering
Neutrino Recoil Spectra\\
\vspace{3mm}
\includegraphics[width=0.48\textwidth]{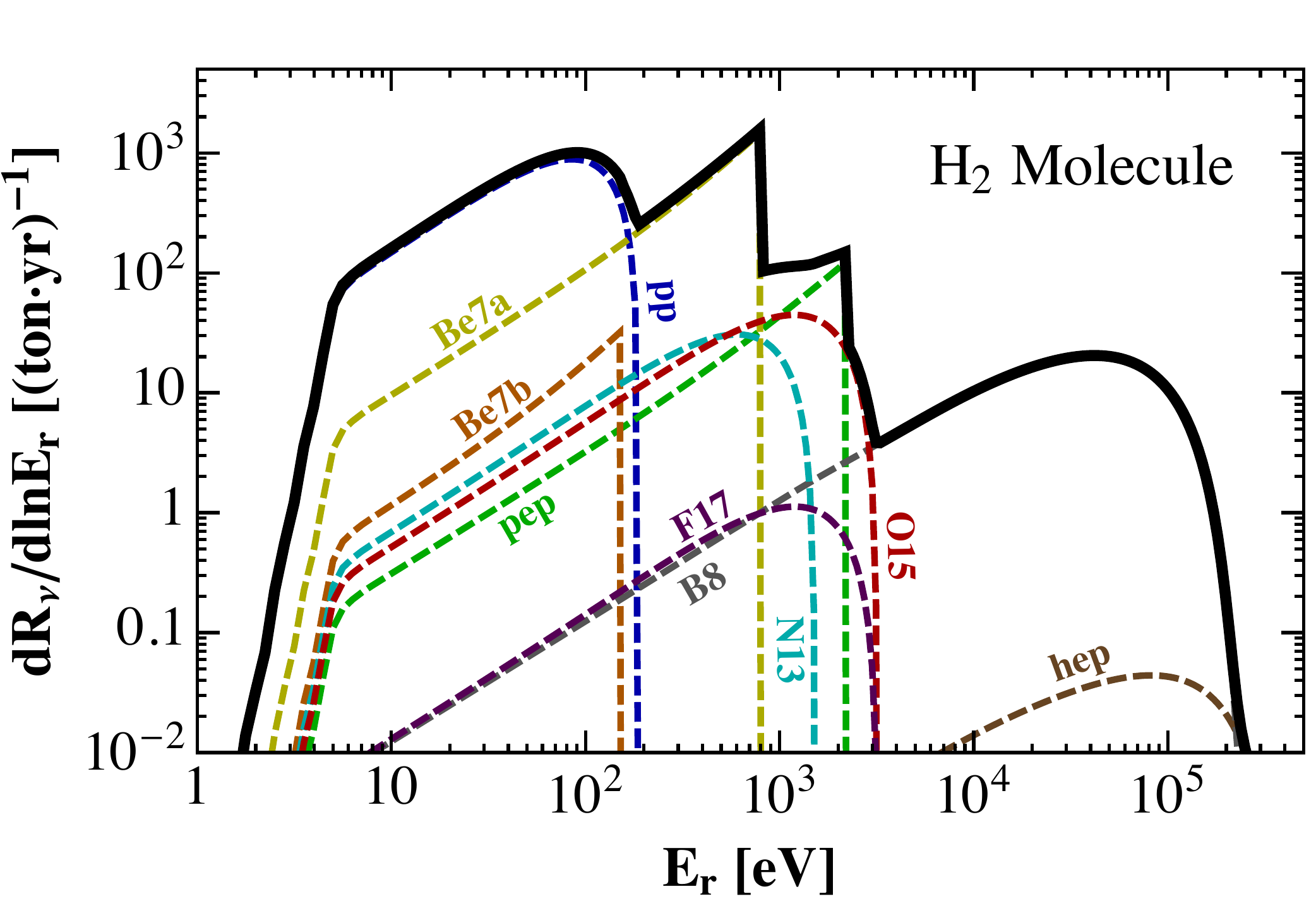}
\hspace{0.3cm}
\includegraphics[width=0.48\textwidth]{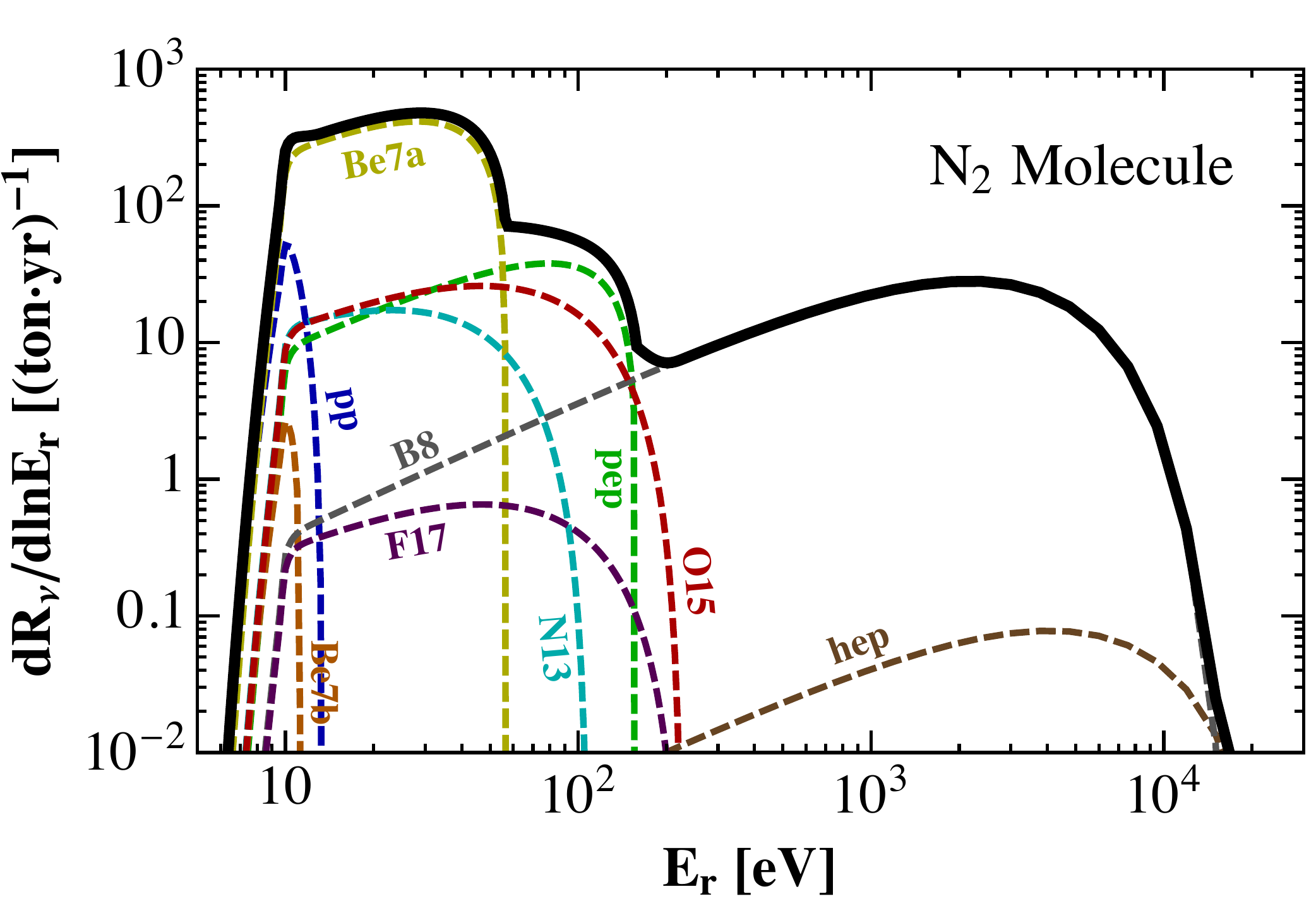}\\
\begin{minipage}[t]{0.5\textwidth}
\mbox{}\\[-0.1\baselineskip]
~~~~~~~\includegraphics[width=0.96\textwidth]{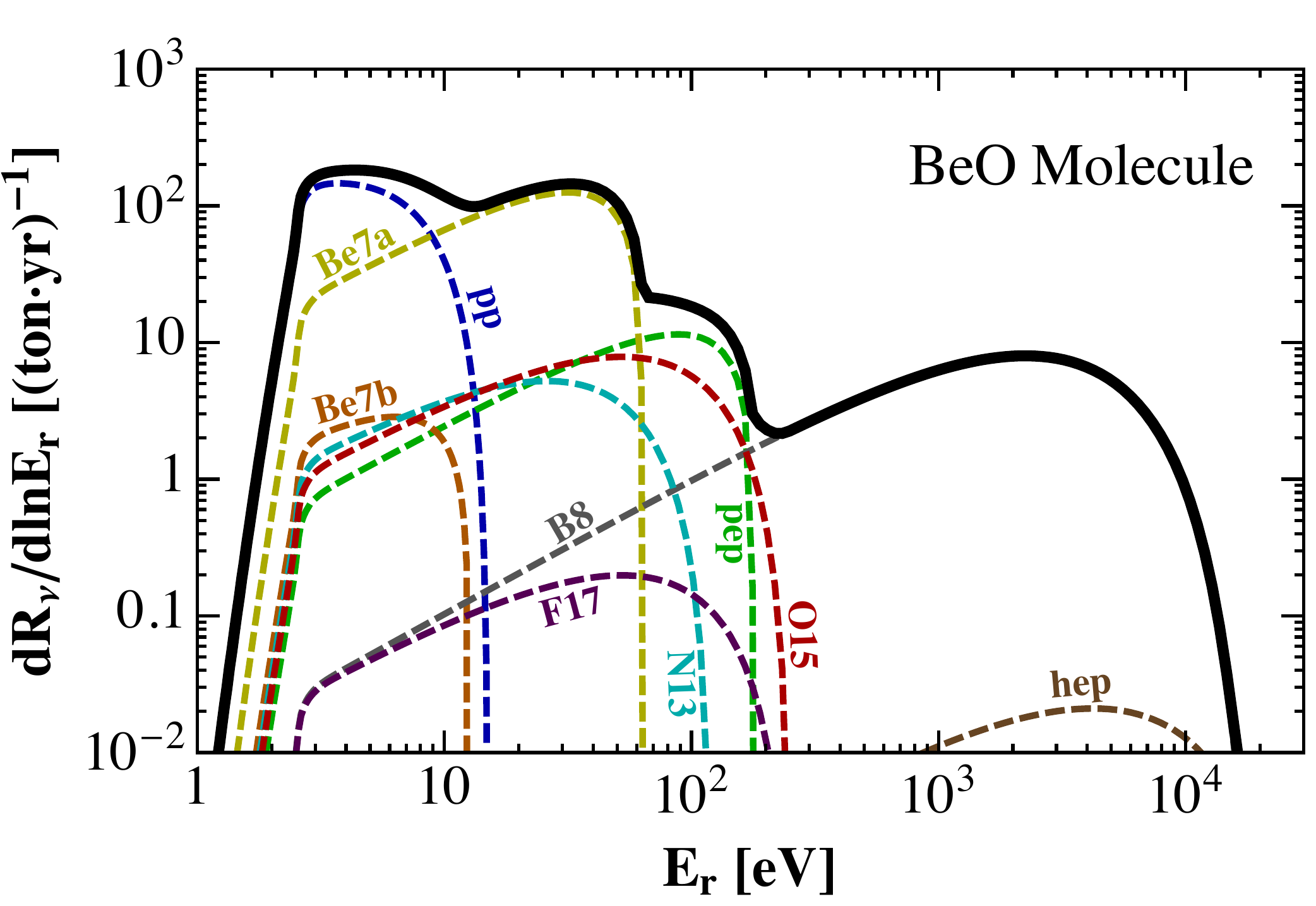} 
\end{minipage}
\hfill
\begin{minipage}[t]{\textwidth}
\mbox{}\\[-\baselineskip]
\caption{
The differential dissociation rate due to solar neutrinos for three example molecules:
H$_2$ with $A=1$ and $\Delta E_B = 4.75 \text{ eV}$ ({\it top left}), N$_2$ with $A=14$  and $\Delta E_B = 9.79 \text{ eV}$  ({\it top right})  and  BeO (interactions with Be only) with $A=9$ for Be and $\Delta E_B = 4.54 \text{ eV}$ ({\it bottom}). The pp neutrinos are pronounced in the H$_2$ and BeO spectra because of the nuclei's relatively low mass and binding energy with respect to N$_2$. The H$_2$ cross section is partly dominated by the spin dependent interaction, which explains the less smooth shape of the spectrum. The total expected rate for H$_2$ is $\sim3040$~events/ton/year, for N$_2$ is $\sim780$~events/ton/year, 
and for BeO is~$\sim480$~events/ton/year.  
\label{fig:dRnu_dEr}
}
\end{minipage}
\end{figure}

The inelastic spin-independent\footnote{The spin-independent neutrino-nucleus cross section is a good approximation for atoms with large atomic numbers. Results shown for hydrogen have been calculated using both the spin-dependent and spin-independent contributions. Details are presented in Appendix~\ref{app:Neutrino_Scattering}.} neutrino-nucleus cross section is,
\beq
\frac{\partial\sigma(E_r,E_\nu)}{\partial E_r} = G^2 \frac{M}{8\pi} [Z(4 \sin^2\theta_W-1) + N]^2 \left(1-\frac{ME_r}{2E_\nu^2}\right) \int d\ln{E_{\rm int}} \FdisE \,.
\eeq
The recoil energy, $E_r = \frac{q^2}{2 M}$, is that of the entire molecule with $M$ denoting the molecular mass and $\FdisE$ is the molecular form factor. The differential dissociation rate is 
\beq
\frac{dR_\nu}{dE_r} = N_T \int\frac{\partial F}{\partial E_\nu}\frac{\partial\sigma(E_r, E_\nu)}{\partial E_r} dE_\nu\,,
\eeq
where $E_\nu$ is the incoming neutrino energy, $\frac{\partial F}{\partial E_\nu}$ is the differential neutrino flux, and $N_T$ is the number of targets in the system.  

Two examples of the differential molecular dissociation rate for neutrino scattering are shown in Fig.~\ref{fig:dRnu_dEr}. Presented are the expected recoil spectra from the solar neutrino flux for 1 ton-year of a hydrogen, nitrogen, and beryllium-oxide molecular target (where the target atom is taken to be Be). Our spectra account for nine families of solar neutrino fluxes, namely: pp, pep, hep, Be7a, Be7b, B8, N13, O15, and F17~\cite{Bahcall:2004pz}. These are the dominant sources of known background for recoil energies expected from a light DM search (for large exposures), while the energies of atmospheric neutrinos are too high. The spectra have some dependance on the target particle. For example, the pp-neutrino contribution is much more prominent in the beryllium oxide spectrum, since beryllium is lighter than nitrogen and the beryllium-oxide binding energy is much smaller than the nitrogen binding energy. The specific shape of the spectrum governs the ability of an experiment to perform background reduction of neutrinos in order to detect a DM signal. The details of such background reduction are discussed below.

\subsection{Neutrino Background Reduction and the Neutrino Floor}
The neutrino spectrum introduces an irreducible background to any DM direct detection experiment. At low exposures, 
the background is negligible.  
However, as the exposure of an experiment increases, this background must be removed in order to detect a signal.

Following~\cite{Ruppin:2014bra}, we have performed a binned maximum-likelihood test-statistics analysis in order to determine the significance of a potential anomaly in the measured spectra. We define the likelihood function as 
\beq
\mathcal{L}(\sigma_{\chi-n},\phi_\nu) = \prod_{i=1}^{N_{bin}}P\left(N^i \biggr\vert \mu_\chi^i+\sum_{j=1}^{N_\nu}\mu_\nu^{i,j}\right) \times \prod_{j=1}^{N_\nu}\mathcal{L}_\nu^j(\phi_\nu^j)\,,
\eeq
where $P(N,\mu)$ is a Poisson distribution, $N_i$ is the number of events in the $i$-th bin, and $\mu_\chi^i, \mu_\nu^{i,j}$ are, respectively, the expected number of events for the DM and neutrino $j$-th  family in the $i$-th bin. The functions $\mathcal{L}^j(\phi_\nu^j)$ are likelihood functions related to the flux normalization of the $j$-th neutrino family. These functions are taken to be normalized Gaussian distributions with standard deviations corresponding to the uncertainty of each neutrino family flux~\cite{Bahcall:2004mq}. A test statistics ratio is defined as follows,
\beq
q_0=
\begin{cases} 
      -2\ln{\left[\frac{\mathcal{L}(\sigma_{\chi-n}=0,\hat{\hat{\phi}}_\nu^j)}{\mathcal{L}(\hat{\sigma}_{\chi-n},\hat{\phi}_\nu^j)}\right]} & \hat{\sigma}_{\chi-n} > 0 \\
      0 & \hat{\sigma}_{\chi-n} < 0 \\
\end{cases}\,,
\label{eq:q0}
\eeq
where values with the carot symbol are treated as nuisance parameters whose values are chosen such that $\mathcal{L}$ is maximal. For a given measurement of $N_i$, a large value of $q_0$ implies a large discrepancy between a neutrino only hypothesis and a neutrino + DM hypothesis.  Following Wilk's theorem \cite{Cowan:2010js}, the function $q_0$ follows a half-$\chi^2$ distribution with one degree of freedom. Thus, the significance of a given value of $q_0$ in units of standard deviations is $Z=\sqrt{q_0}$.

To understand the impact of neutrinos on the reach of an experiment, we assume now an idealized experiment that has no backgrounds except for, possibly, neutrinos. For exposures low enough such that the neutrino background is effectively zero, the cross section reach scales simply as $1/MT$, where $MT$ denotes the exposure (target mass $\times$ time).  As neutrino events begin to enter the signal region, their uncertainty reduces the significance of a DM signal thereby weakening the cross section reach. This uncertainty has two main contributions. The first is a Poisson contribution originating from the Poisson behavior of the number of neutrino events. The second contribution is the uncertainty in the solar neutrino flux.  This uncertainty is assumed to be gaussian distributed. If the neutrino and DM spectra are significantly different from each other, and if the experiment has sufficient energy resolution, then this extra handle can reduce the effect of the neutrino uncertainty. However, if these requirements are not fulfilled, the overall cross-section reach scales as the inverse of the neutrino uncertainty over the total number of signal events. The evolution of the cross section reach can be understood by considering a DM signal that perfectly mimics some part of the neutrino spectrum. In this case, the reach is proportional to~\cite{Ruppin:2014bra, Billard:2013qya} 
\beq
\sigma_{\text{reach}} \propto \frac{\sqrt{MT+\epsilon^2(MT)^2}}{MT}\,,
\eeq
where $\epsilon$ is the uncertainty of some member of the neutrino flux. When $\epsilon^2MT\ll1$ the uncertainty is in the Poisson regime, and the cross section reach scales as $1/\sqrt{MT}$. At larger exposures when $\epsilon^2 MT\gg1$, the flux uncertainty dominates, and the cross section reach saturates and becomes constant with growing exposure. If the exposure is further increased, the tail of the neutrino spectrum becomes measurable, and this additional discriminating power allows the experiment to return to the Poisson regime where again the reach scales as $1/\sqrt{MT}$.

For a given DM mass, there is an optimal exposure and threshold that allows for the largest cross section reach before reaching the saturation regime.\footnote{For some DM masses, the spectrum is sufficiently different from the neutrino spectrum such that the saturation regime may be totally avoided.} This defines the so-called neutrino floor. In this study, we have calculated the cross section reach by finding the parameters for which the test statistics defined in Eq.~(\ref{eq:q0}) gives a $2\sigma$ significance. This has been done for a number of exposures and for a recoil-energy threshold corresponding to the quantum momentum transfer threshold discussed in Sec.~\ref{sec:quantum-effects}. Setting the threshold to be the classical threshold changes the sensitivity only for the lightest DM masses.

The results are presented in Fig.~\ref{Fig:Sens_Plots} for an exposure of 0.1 ton$\cdot$years. For all three molecules, the Be7a, O15, and B8 uncertainties limit  dominate the total neutrino flux uncertainty. The pp and pep components contribute to the total spectrum, but have much smaller uncertainties. All other components have a negligible contribution to the neutrino background for all three molecules. The features in the cross section reach correspond to DM masses whose spectra happen to mimic a component of the neutrino spectrum. The various components are marked on the figures, except for cases where the DM spectra are significantly different from the neutrino spectra for all DM masses. In the latter case, the dominant uncertainty is just the Poisson uncertainty and no features are present.

%%%%%%%%%%%%%%%%%%%%%%%%%%%
\section{Discussion: Towards a Real Experimental Setups} 
\label{sec:setups}
%%%%%%%%%%%%%%%%%%%%%%%%%%%

In the  study presented above, we do not attempt to describe a realistic experimental setup that allows for the measurement of bond-breaking events. Instead, the aim of this paper is to set up the problem and study the prospects of an ideal experiment that would be sensitive to such low-energy events.  Of course, a true experiment could suffer from numerous backgrounds and will have various experimental challenges that will require a dedicated study. While this is beyond the scope of this paper, there are interesting features that would motivate the use of crystals for the detection of bond-breaking events. In particular, in addition to having relatively low binding energies, crystals may allow one to enhance the signal spectroscopically  and also to discriminate between low-energy (signal) and high-energy (background) events as well as electron versus nuclear recoils. A more detailed study will appear in upcoming publications~\cite{color-center, color-center-theory}.

As mentioned above, the choice of target material is crucial. Crystals composed of ions that are heavy, allow for an enhanced signal due to the coherence effect, while those composed of light ions allow for sensitivity to lower DM masses. One could potentially choose a target crystal that is composed of both light and heavy nuclei, thus utilizing the advantages of both and gaining sensitivity to a wide range of DM masses and cross sections. As discussed in Sec.~\ref{sec:LightDM} and derived in App.~\ref{app:optimization}, the expected signal rate depends on the mass number of the scattered atom, $A$ (or equivalently $m_1$), the binding energy $\Delta E_B$, and the DM mass, $m_\chi$.  For instance, for a trivial DM form factor, the maximum rate behaves as 
\beq
R_{\rm max} \propto \left\{ 
\begin{tabular}{lcc}
$A^{1/2} \cdot \DE^{-1/2}$ && $m_\chi \ll m_n,m_1$
\\
 $A^3 \cdot m_{\chi}^{-1}$ && $m_\chi \gg m_n,m_1$ 
 \end{tabular}\right. \,.
\eeq
The above results can be seen in Fig.~\ref{Fig:Sens_Plots}, where dissociation rates have been plotted for diatomic molecules with varying elements and binding energies. For the case of $\FDM=1$, in order to increase the maximal rate it is better to have large $A$ and small $\DE$. On the other hand, the mass threshold (which depends on $q_{\rm min}$) is always lower for smaller $A$ and $\DE$ as can be seen from Eqs.~\eqref{eq:Class_Kinematics_1}, \eqref{eq:Quant_qmin}, and \eqref{eq:m_Threshold}. For more details see App.~\ref{app:optimization}. With the above in mind, a target material may be chosen in order to maximize sensitivity to a given feebly-interacting particle, be it light dark matter or a specific neutrino spectrum.

To summarize, while much R\&D is still required to employ bond-breaking interactions for the detection of weakly interacting particles, the potential is significant.  This paper takes a first step towards exploring such directions. With growing theoretical motivation for light dark matter in the sub-GeV mass range, and very few current experimental capabilities to explore them, we find the prospect of bond-breaking searches very appealing and strongly encourage further studies of possible experimental setups that may realize this in practice.

%%%%%%%%%%%%%%%%%%%%%%%%%%%%%%%%%%%%%%%%%%%%%%%%%%%%%%%%%%%%%%%%%
\section*{Acknowledgement}
We thank Ranny Budnik and Ori Cheshnovsky for many useful discussions. 
R.E.~is supported by the DoE Early Career research program DESC0008061 and through a Sloan Foundation Research Fellowship. O.S. is supported in part by a grant from the Clore Israel Foundation. T.V. is supported by the European Research Council (ERC) under the EU Horizon 2020 Programme (ERC-CoG-2015 - Proposal n. 682676 LDMThExp), by the PAZI foundation, by the German-Israeli Foundation (grant No. I-1283- 303.7/2014) and by the I-CORE Program of the Planning Budgeting Committee and the Israel Science Foundation (grant No. 1937/12). J.M. was supported by grant DE-SC0012012. 

%%%%%%%%%%%%%%%%%%%%%%%%%%%%%%%%%%%%%%%%%%%%%%%%%%%%%%%%%%%%%%%%%
%%%%%%%%%%%%%%%%%%%%%%%%%%%%%%%%%%%%%%%%%%%%%%%%%%%%%%%%%%%%%%%%%
%%%%%%%%%%%%%%%%%%%%%%%%%%%%%%%%%%%%%%%%%%%%%%%%%%%%%%%%%%%%%%%%%
\appendix

\section{Derivation of the Form Factor}
\label{Appendix:FF}
In this appendix, we study the evaluation of the form factor, deriving both the exact solution in the spherically symmetric case, and the form factor in the Born approximation. Our starting point is Eq.~(\ref{eq:Fdis}), which we reproduce here:
\beq
\Fdis = \frac{\qt^3}{(2\pi)^3}\int d\Ot \left| \int d^3r e^{i\mum\qv\cdot\rv} \Psi_{\qtv}^*(\rv) \Psi_i(\rv) \right|^2\,.
\label{eq:FF_def}
\eeq
\subsection{Evaluation of the Exact Solution}
\label{Appendix:Full}
Calculating the exact form factor involves solving for the eigenfunctions of the unbound Schr\"odinger equation.  This can be done numerically by summing up solutions for all values of angular momentum and for each value of internal recoil energy, $E_{\rm int}$.  The final wavefunctions can be expanded in terms of spherical harmonics,
\beq
\Psi_{\qtv}(\rv) = 4 \pi \sum_{\ell,m} \frac{a_{\ell}}{2\ell+1} \Ylm_{\ell m}^*(\Omega_{\qt}) \Ylm_{\ell m}(\Omega_r) R_{\qt \ell}(r)\,,
\label{eq:Psi_Sph_Hrm}
\eeq
where $\Omega_{\qt}$, $\Omega_r$ are solid angles of the vectors $\qtv$ and $\rv$ with respect to $\qv$, respectively. The exponent can be expanded as,
\beq
e^{i\mum\qv\cdot\rv} = 4 \pi \sum_{\ell,m} i^{\ell} \Ylm_{\ell m}^*(\Omega_{q}) \Ylm_{\ell m}(\Omega_r) j_{\ell}\Big(\mum q r\Big)\,,
\label{eq:Exp_Sph_Hrm}
\eeq
where $j_{\ell}$ are spherical Bessel functions. Finally, the absolute value in Eq.~(\ref{eq:FF_def}) can be rewritten as,
\beq
\left| \int d^3r e^{i\mum\qv\cdot\rv} \Psi_{\qtv}^*(\rv) \Psi_i(\rv) \right|^2 = \left[ \int d^3r' e^{i\mum\qv\cdot\rv'} \Psi_{\qtv}^*(\rv') \Psi_i(\rv') \right] \left[ \int d^3r'' e^{-i\mum\qv\cdot\rv''} \Psi_{\qtv}(\rv'') \Psi_i^*(\rv'') \right]\,.
\label{eq:abs_val}
\eeq
Substituting Eqs.~\eqref{eq:Psi_Sph_Hrm}-\eqref{eq:abs_val} into Eq.~(\ref{eq:FF_def}), integrating over $\Omega_{\qt}\,,\Omega_{r'},\Omega_{r''}$ for a spherically-symmetric initial state, and using the identity 
\beq
\sum_{m=-\ell}^\ell \Ylm_{\ell m}^*(\Omega) \Ylm_{\ell m}(\Omega) = \frac{2\ell +1}{4\pi}\,,
\eeq
the full expression for the form factor as a function of the eigenstates of the unbound wavefuntion becomes 
\beq
\Fdisq = 8 \qt^3 \sum_{\ell} \frac{|a_\ell|^2}{(2\ell+1)} \left| \int dr r^2 j_{\ell}\Big(\mum qr\Big) R_{\qt \ell}(r) \Psi_i(r) \right|^2\,.
\label{eq:Fdis_Gen}
\eeq
If the final wavefunctions are plane waves at large distances from the origin then it follows that Eqs.~\eqref{eq:Psi_Sph_Hrm} and \eqref{eq:Exp_Sph_Hrm} are equivalent asymptotically. In this case, $a_\ell = i^\ell (2 \ell +1)$.  Under this assumption, and choosing a spherically-symmetric ground state for the initial wavefunction,
\beq
\Psi_{i,n\ell m} = \Psi_{i,000} = \Y_{00} R_{00}(r)\,,
\eeq
and Eq.~(\ref{eq:Fdis_Gen}) becomes 
\beq
\Fdisq = \frac{2}{\pi} \qt^3 \sum_{\ell} (2\ell+1) \left| \int dr r^2 j_{\ell}\Big(\mum qr\Big) R_{\qt \ell}(r) R_{00}(r) \right|^2\,.
\label{eq:Fdis_Ground}
\eeq

\subsection{The Born Approximation}
\label{Appendix:Born}
The Born approximation of the form factor in Eq.~\eqref{eq:FF_def} is obtained by taking the initial bound wavefunction to be the solution of the 
Schr\"odinger equation, solved for the Morse potential, and by assuming plane waves for the radial solution of the final state,
\beq
\Fdisq = \frac{\qt^3}{(2\pi)^3} \int d\Ot \left| \int d^3r e^{i\mum\qv\cdot\rv} e^{-i\qtv\cdot\rv} \Psi_i(\rv) \right|^2\,.
\label{eq:Fdis_App}
\eeq
We define,
\beq
\Kv \equiv - \left( \mum \qv - \qtv \right)\,,
\eeq
and 
\beq
K^2 = \qt^2 + \left(\mum\right)^2 q^2 - 2 \mum q \qt c_{\theta\qt}\,,
\eeq
with $c_{\theta\qt}$ the cosine of the angle between the vectors $\qv$ and $\qtv$.
For an initial state that is symmetric around the direction of $\qv$, we then obtain  
\beq
\Fdisq = \frac{m_1}{\mu_{12}} \frac{\qt^2}{(2\pi)^2} \int_{\qt-\mum q}^{\qt+\mum q} \frac{K dK}{q} \left| \int d^3r e^{-i\Kv\cdot\rv} \Psi_i(\rv) \right|^2.
\eeq
For $\Psi_i(\rv) = Y_{00} R_{00}(r)$,
\begin{eqnarray}
\Fdisq		& = &	\frac{m_1}{\mu_{12}} \frac{\qt^2}{4\pi} \int_{\qt-\mum q}^{\qt+\mum q} \frac{K dK}{q} \left| \int dc_\theta drr^2 e^{-iKrc_\theta} R_{00}(r) \right|^2 \\
				& = &	\frac{m_1}{\mu_{12}} \frac{\qt^2}{\pi q} \int_{\qt-\mum q}^{\qt+\mum q} \frac{dK}{K} \left| \int drr \sin(Kr) R_{00}(r) \right|^2\,.
\end{eqnarray}
Approximating the initial radial wave function with a Gaussian of the form,
\beq
R_{00}(r) = \left(\frac{1}{\sigma_{0}\sqrt{\pi}}\right)^{\frac{1}{2}} \frac{e^{-\frac{1}{2}\left(\frac{r-r_{0}}{\sigma_{0}}\right)^{2}}}{r}\,,
\eeq
the Born approximation for the form factor takes the form,
\beq
\Fdisq = \frac{m_1}{\mu_{12}} \frac{\qt^2}{\pi^{3/2} \sigma_{0} q} \int_{\qt-\mum q}^{\qt+\mum q} \frac{dK}{K} \left| \int dr \sin(Kr) e^{-\frac{1}{2}\left(\frac{r-r_{0}}{\sigma_{0}}\right)^{2}} \right|^2\,.
\label{eq:Fdisq_app}
\eeq

\section{Neutrino Scattering Rate}
\label{app:Neutrino_Scattering}

The differential neutrino-molecule cross section is given by,
\begin{eqnarray}
\frac{\partial \sigma(E_r,E_\nu)}{\partial E_r} & = & G^2 \frac{M}{2\pi} \left[ (G_V+G_A)^2 + (G_V-G_A)^2\left(1-\frac{E_r}{E_\nu}\right)^2 - (G_V^2-G_A^2)\frac{M E_r}{E_\nu^2} \right] \nonumber \\
& & \times \int d\ln{E_{\rm int}} \FdisE  \,,
\end{eqnarray}
where $G$ is the Fermi constant and $G_V,G_A$ are the vector and axial contributions to the hadronic current, respectively. They can be parameterized as,
\begin{eqnarray}
G_V & \equiv & \left[g_V^p Z + g_V^n N \right] F_{\rm nuc}^V(q^2)\,, \nonumber \\
G_A & \equiv & \left[g_A^p (Z_+ - Z_-) + g_A^n (N_+ - N_-) \right] F_{\rm nuc}^A(q^2)  \,,
\end{eqnarray}
where $Z$ ($N$) is the number of protons (neutrons) and $Z_{\pm}$ ($N_{\pm}$) are the number of protons (neutrons) with spin plus or minus. For our purposes, the form factors  $F_{\rm nuc}^{V/A}(q^2)$ are just unity, since the momentum transfer is small. The appropriate values of the coefficients $g_V^p$, $g_V^n$, $g_A^p$, and $g_A^n$ can be found in \cite{Freedman:1977xn}.

For large nuclei, the spin dependent (SD) part of the cross section is negligible and we obtain,
\beq
\frac{\partial \sigma(E_r,E_\nu)}{dE_r} \simeq G^2 \frac{M}{8\pi} [Z(4 \sin^2\theta_W-1) + N]^2 \left(1-\frac{ME_r}{2E_\nu^2}\right) \int d\ln{E_{\rm int}} \FdisE\,.
\eeq

For small nuclei like hydrogen, the SD cross section cannot be neglected. For molecular hydrogen,  we obtain 
\beq
\frac{\partial \sigma(E_r,E_\nu)}{\partial E_r} \simeq G^2 \frac{M}{8\pi} \times 1.46 \left[1 + \frac{mE_r}{2 E_\nu^2}\right] \int d\ln{E_{int}} \FdisE\,.
\eeq

\section{Optimization of Target Material}
\label{app:optimization}

Various considerations will affect the choice of the target material for a realistic setup. Larger nuclei will have larger cross sections, because of coherence, but also a suppressed energy transfer, because of large masses, and less targets per kg of detector material. Smaller binding energies will allow for larger interaction rates and lower mass thresholds. In the classical regime, and for a given DM-nucleon cross section, the dissociation rate for scattering events with atom type $i$ within a given molecule scales as 
\beq
R_i \propto \left[ f_PZ_i+f_N(A_i-Z_i) \right]^2 \frac{N_{Ti}}{m_\chi \mu_{\chi n}^2} \int dq^2 \times \FDM \Theta\left(q - \sqrt{\frac{2 m_i^2}{\mu_{12}} \DE}\right) \eta(v_{\rm min, i})\,.
\eeq
In what follows, we present the approximate scaling of the expected interaction rate for two regimes of DM masses and as a function of the target parameters.

For a given total target mass $M_{tot}$ and taking $f_P = f_N$, the rate scales as,
\beq
R_i \propto \frac{A_i^2}{m_{\chi} \mu_{\chi n}^2} \frac{M_{tot}}{m_1+m_2} \int_{q_{min,i}}^{q_{max,i}} dq^2 \times \FDM\,,
\label{eq:Rate_Clas_approx}
\eeq
where we have used $N_{Ti} = \frac{M_{tot}}{m_1+m_2}$ and $m_2$ is the mass of second atom within the diatomic molecule. The classical minimal and maximal values of $q$ are given by,
\begin{eqnarray}
q_{min,i}^2 & = & 2(m_1+m_2) \frac{m_i}{m_2} \DE \,,\nonumber \\
q_{max,i}^2 & = & 4 \mu_{\chi i}^2 v_{max}^2\,. \
\end{eqnarray}
For simplicity we take $m_1 = m_2$. Then the variables in Eq.~(\ref{eq:Rate_Clas_approx}) scale as 
\beq
A \propto m_1 \propto N_T^{-1}\,,
\eeq
where we have suppressed the $i$ index. At the value of $m_\chi$ for which the rate peaks just above the threshold mass, one finds $m_{\chi}^2 \propto m_1 \DE$ (as long as $m_\chi \ll m_1$).  (For the case where $m_1 \neq m_2$ these relations are slightly modified.)

For $\FDM=1$, 
\beq
R \propto \frac{A^2}{m_{\chi} \mu_{\chi n}^2 m_1} \left( q_{max}^2 - q_{min}^2 \right)\,,
\eeq
while for $\FDM = \left( \frac{q_0}{q} \right)^4$\,,
\beq
R \propto \frac{A^2}{m_{\chi} \mu_{\chi n}^2 m_1} \left( \frac{1}{q_{min}^2} - \frac{1}{q_{max}^2} \right)\,.
\eeq
Taking the limits $m_\chi \ll m_n$ and $m_\chi \gg m_n$, and choosing the maximal rate in each regime, the results (up to $\mathcal{O}(1)$ factors) is,
\beq
R_{\rm max} \propto
\begin{tabular}{| l | c | c |}
\hline
& $m_\chi \ll m_n,m_1$ & $m_\chi \gg m_n,m_1$ \\
\hline
$\FDM = 1$ & $A^{1/2} \cdot \DE^{-1/2}$ & $A^3 \cdot m_{\chi}^{-1}$ \\
\hline
$\FDM = \left( q_0/q \right)^4 $ & $A^{-3/2} \cdot \DE^{-5/2}$ & $\DE^{-1} \cdot m_\chi^{-1}$ \\
\hline
\end{tabular}\,.
\eeq

The above results can be seen in Fig.~\ref{Fig:Sens_Plots}, where dissociation rates are shown for diatomic molecules for various elements and binding energies. Small binding energies always increase the dissociation rate, except for $\FDM=1$ and large DM mass, for which the binding-energy dependence vanishes. For $\FDM \propto 1/q^4$, the $A$ dependence vanishes for large DM masses.  
In the low DM mass regime, the maximal rate's dependence on $A$ and $\DE$ highly depends on the DM form factor. 
For $\FDM=1$, it is better to have large $A$ and small $\DE$, while for the non-trivial form factor, it is better to have small $A$ 
and small $\DE$.  

\bibliographystyle{ieeetr}
\bibliography{MolecularDetection}

\end{document}